\documentclass[a4paper,11pt]{article}

\usepackage{a4wide}
\usepackage[latin1]{inputenc}
\usepackage[T1]{fontenc}
\usepackage[english]{babel}
\usepackage{graphicx,subfigure}
\usepackage{textcomp}
\usepackage{geometry}
\usepackage{bm}
\usepackage{setspace}
\usepackage{color}
\usepackage{latexsym}
\usepackage{amssymb}

\usepackage[numbers]{natbib}

\begin{document}

\begin{center}
\LARGE{Separation between coherent and turbulent fluctuations. What can we learn from
the Empirical Mode Decomposition?}
\end{center}

\begin{center}
Nicolas Mazellier\footnote{Corresponding author: nicolas.mazellier@univ-orleans.fr} and Fabrice Foucher
\end{center}


\begin{center}
\small{Institut PRISME, 8, rue L{\'e}onard de Vinci, 45072 Orl{\'e}ans, FRANCE}
\end{center}


\centerline{\bf Abstract}

\noindent
The performances of a new data processing technique, namely the Empirical Mode
Decomposition, are evaluated on a fully developed turbulent velocity signal
perturbed by a numerical forcing which mimics a long-period flapping.
First, we introduce a "resemblance"
criterion to discriminate between the polluted and the unpolluted modes extracted
from the perturbed velocity signal by means of the Empirical Mode Decomposition
algorithm. A rejection procedure, playing, somehow, the role of a high-pass filter,
is then designed in order to infer the original velocity signal from the perturbed one.
The quality of this recovering procedure is extensively evaluated in the case of
a "mono-component" perturbation (sine wave) by varying both the amplitude and the frequency
of the perturbation. An excellent agreement between the recovered and the reference
velocity signals is found, even though some discrepancies are observed when the
perturbation frequency overlaps the frequency range corresponding to the energy-containing
eddies as emphasized by both the energy spectrum and the structure functions.
Finally, our recovering procedure is successfully performed on a
time-dependent perturbation (linear chirp) covering a broad range of frequencies.

\vskip 1truecm

\section{Introduction}
\label{intro}

Most, if not all, engineering applications involving turbulent fluid flow undergo
the influence of the so-called coherent structures. Even though their exact definition
is still open to debate \cite{Hussain1983}, their existence in turbulent flows is now
commonly recognized since the beautiful flow visualization of a turbulent mixing layer
by Brown and Roshko \cite{BrownRoshko1974}. Their exact role in the onset of turbulence and their
impact on the dynamics of turbulent interactions are of major interest in the development
of relevant turbulence models. In particular, the strong persistence in space
and time of such coherent structures may induce long-range memory implying, among
others, a strong dependence of turbulence on boundary and initial conditions (see
e.g. \cite{George1992}). Evidence of such interplay were reported in recent experiments
led, for instance, by Tong and Warhaft \cite{TongWarhaft1994} on jet modification by means of a 
small ring located in the initial shear-layers, by Hu et al. \cite{Huetal2006} who investigated
the effect of the corner radius onto the near-wake of a square cylinder or by
Lavoie et al. \cite{Lavoieetal2005} who studied the onset of turbulence generated by various
shape of regular grid. Besides the fact that the flow topology is extremely
sensitive to the modification of the initial conditions, the fundamental issue of their
influence onto the energy transfer is still misunderstood \cite{Hussain1983}.
This issue relates intimately to the concept of universality which is a key point in
turbulence modeling.

Besides the fundamental issue of the interplay between coherent structures and turbulence,
their management or control has attracted an increasingly interest in the scientific
community in order to either favor or prevent their effects on a given process. For instance,
the enhancement of mixing, required in processes such as non-premixed combustion, chemical
reaction or wall cooling, can be achieved through the joint action of large-scale convection
by coherent structures and the micro-mixing induced by turbulence. On the contrary,
noise generation \cite{TongWarhaft1994} and fluid-structure interaction phenomena, such
as vortex-induced vibration experienced by freely movable bluff-body (see
\cite{Bearman1984}),
are extremely sensitive to the topology of coherent structures. Among the various strategies
of flow control, the active closed-loop concept (referred to as reactive by Gad-el-Hak
\cite{GadElHak2000}) appears as the most attractive \cite{Choietal2008}. This approach is based
on a feedback loop allowing for a real-time adaptation of the actuation in order to optimize a
net energy gain. Unfortunately, unlike laminar flows, the straightforward implementation of
this control strategy in turbulent flows is often disappointing and weakly efficient. The main
drawback relies on the reliability of the reference signal used to drive to feedback loop control.
Indeed, the pollution of this signal, which is basically expected to lock-in with the coherent
structures, by the random fluctuations strongly alters the robustness of the control loop.

For all the above mentioned reasons, intense efforts have been dedicated to develop
data analysis enabling to identify and extract the signature of the coherent structures
from a turbulent signal. Hussain and Reynolds \cite{HussainReynolds1970}
introduced the so-called triple decomposition such that a physical variable
of the turbulent flow, say the velocity for instance, results from the
superimposition of three contributions: a coherent (i.e. phase-averaged),
a mean (i.e. time average) and a random. Even though this definition
allows for a clear separation between coherent structures and turbulent
fluctuations, its application, in practice, is far to be easily tractable.
Other attractive methods have been developed in order to organized motion
in flows. One can cite, for instance, the Proper Orthogonal Decomposition
(see e.g. \cite{Berkoozetal1993} and references therein) which extracts modes
sorted by their energy and is often used to define Reduced-Order Models (ROM)
by keeping the first most energetic modes. Nevertheless, this approach provides
a statistical description and therefore is subjected to convergence criteria 
which restrict its interest in the case of non-stationary dynamics.
Inspired from the multi-scale feature of turbulence, wavelet analysis has
brought significant insights in the understanding of the nonlinear interactions
between the turbulent scales (see e.g. \cite{Fargeetal2001}). The main
drawback of wavelet approach stands in the prescription of a given
class of functions as the decomposition base. 

In the nineties, Huang et al. \cite{Huangetal1998} overcame these drawbacks by introducing
the so-called Empirical Mode Decomposition (EMD). Basically, this data analysis
technique decomposes a signal as the sum of a detail and a trend using its local
envelopes. It is, therefore, entirely driven by the data themselves. In other
words, the EMD does not require any \textit{a priori} base function appearing,
therefore, as an alternative way to investigate both non-stationary and non-linear
physics. Although the EMD encountered a great success in various fields such as
physics or biology, for instance, it has been introduced only very recently in the
framework of turbulence. Huang et al. \cite{Huangetal2008} investigated the intermittency of
fully-developed turbulence by means of EMD and Hilbert transform.
Foucher and Ravier \cite{FoucherRavier2010} used the EMD in order to provide a triple-decomposition
of turbulent signals forced by periodic or random excitation.

The work reported here is dedicated to the assessment of the performances of
the EMD algorithm in discriminating between coherent signature and random
fluctuation. For that purpose, the EMD is applied on a turbulent fluctuation
perturbed by a numerical signal mimicking a long-period flapping. An original
rejection procedure, based on a "resemblance" criterion, is proposed for
discriminating between both contributions. The relevance of the EMD in separating
the turbulent signal from the perturbation is evaluated in numerous conditions
by varying both the amplitude and the frequency of the numerical flapping.  

The paper is organized as follows. The basics of EMD analysis are given
in Sec. \ref{sec:EMD}. Then, the methodology used to separate the perturbation
and the turbulent signal is described in Sec. \ref{sec:method}. Especially, the
"resemblance" criterion and the rejection procedure are introduced. Finally,
the performances of the EMD is analyzed and discussed in Sec. \ref{sec:results}.

\section{The Empirical Mode Decomposition algorithm}
\label{sec:EMD}

In this section, the basics of EMD are briefly reminded and illustrated
via a simple example. More details about the EMD method can be found in
\cite{Huangetal1998}. It is worth noticing that the EMD algorithm used
in this study is computed by means of the Matlab scripts provided by
Flandrin (http://perso.ens-lyon.fr/patrick.flandrin/).

\subsection{The principles}

The EMD method is based on the assumption that a signal $s(t)$ can be
decomposed into the sum of a detail $d(t)$ and a trend $r(t)$, i.e.:

\begin{equation}
s(t) = d(t) + r(t).
\label{eq:EMDbasics}
\end{equation}

The detail $d(t)$ is characterized by high frequency variations, whilst
the trend $r(t)$ represents the low frequency variation. To illustrate this
concept, let us consider the following "two-components" signal

\begin{equation}
s(t) = \sin{2 \pi t} + \frac{1}{2} \sin{20 \pi t}.
\label{eq:ExSignal}
\end{equation}

\begin{figure}[htbp]
\begin{minipage}[c]{.36\linewidth}
\subfigure[]
{\includegraphics[width=1.1\textwidth]{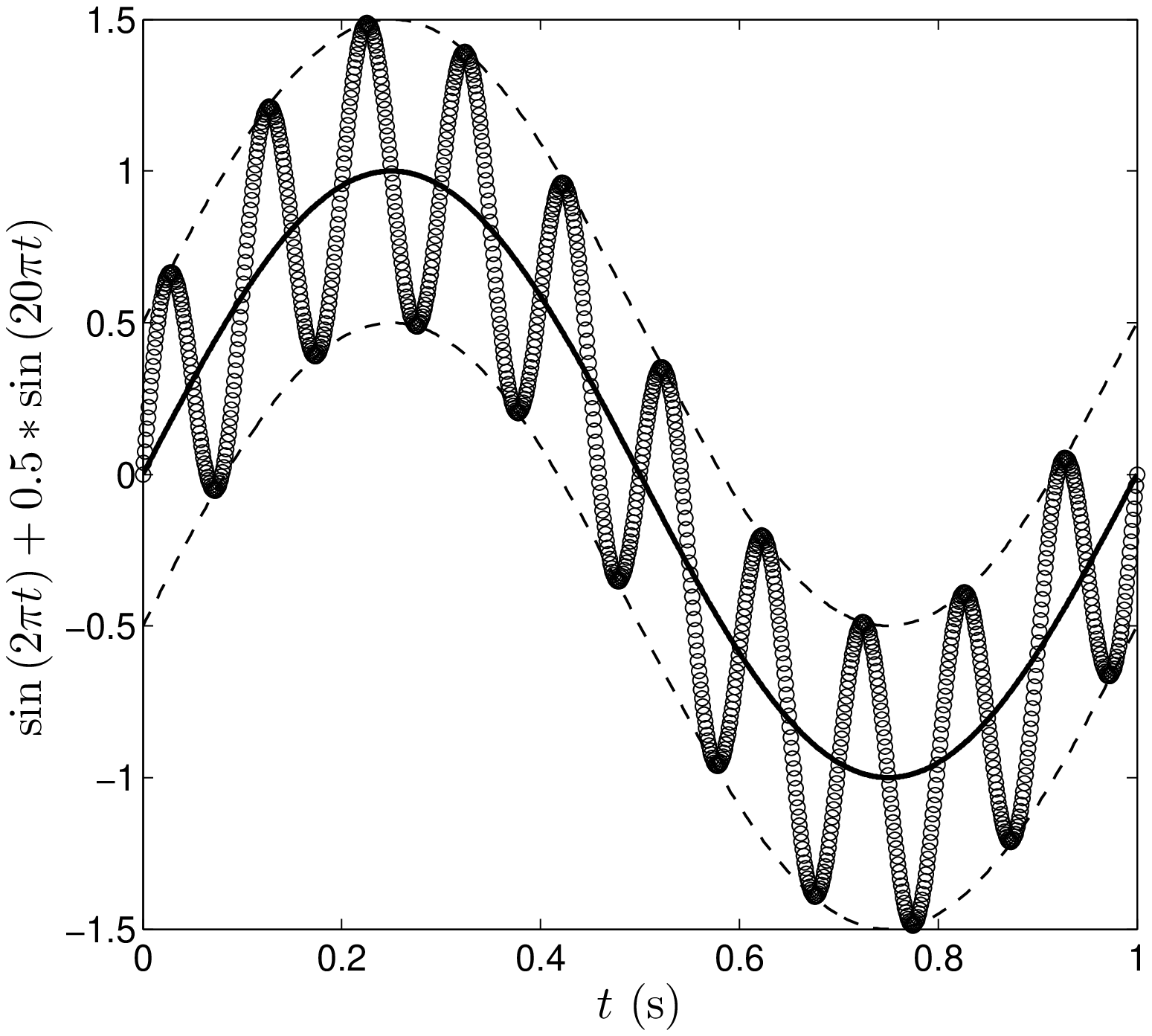}
\label{fig:TwoModes}}
\end{minipage}
\hfill
\begin{minipage}[c]{.56\linewidth}
\subfigure[]
{\includegraphics[width=0.9\textwidth]{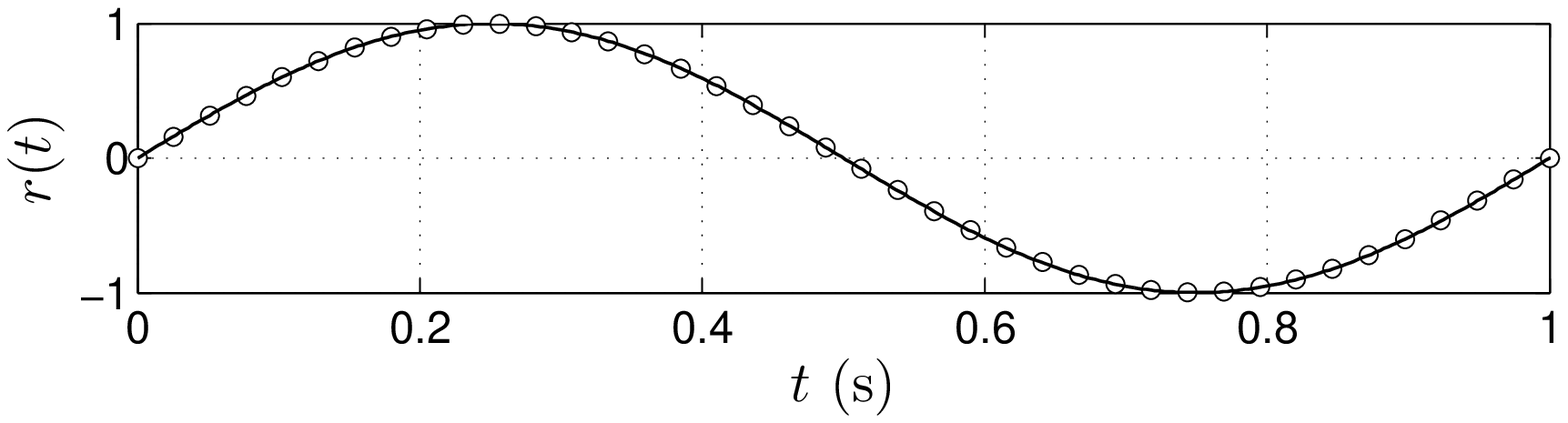}
\label{fig:TwoModesTrend}} 
\subfigure[]
{\includegraphics[width=0.9\textwidth]{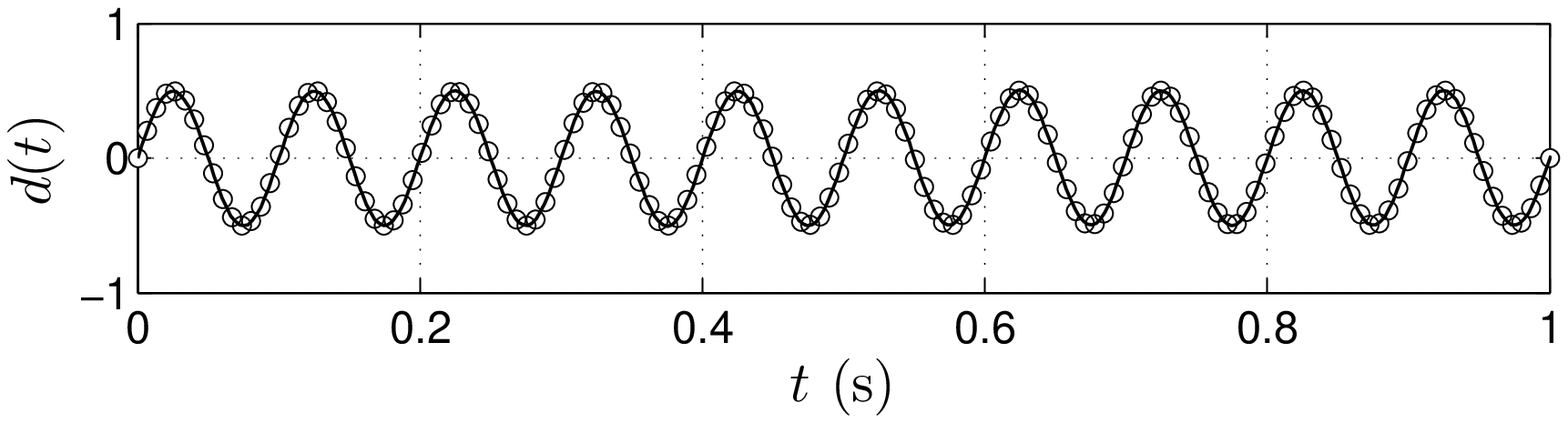}
\label{fig:TwoModesDetail}}
\end{minipage}
\caption{Typical EMD analysis of a "two-components" signal
$s(t) = \sin{2 \pi t} + \frac{1}{2} \sin{20 \pi t}$. (a) Evaluation
of the local mean-value (thick solid line) as the average between the
lower and the upper envelopes (dash lines). (b) Comparison of the
estimated trend (solid line) and the low-frequency component
$\sin{2 \pi t}$ ($\circ$). (c) Comparison of the estimated detail
(solid line) and the high-frequency component
$\frac{1}{2} \sin{20 \pi t}$ ($\circ$).}
\end{figure}

This signal is displayed in Fig. \ref{fig:TwoModes} which clearly evidences
two time-scales. Rilling and Flandrin (2008) reported a complete investigation
of the performances of EMD to analyze similar "two-components"
signals. In particular, they evaluated the sensitivity of the EMD algorithm
when varying separately the frequency ratio and the amplitude ratio between both
contributions. It is worth noting that the specific example we use here belongs
to the range delineated by Rilling and Flandrin (2008) where the EMD succeeds
in separating the two tones.

In order to extract the trend $r(t)$ from the signal,
Huang et al. (1998) introduced the concept of the local mean value
defined as the average between the lower ($e_{min}(t)$) and the upper
($e_{max}(t)$) envelopes of the signal, i.e.

\begin{equation}
r(t) = \frac{e_{min}(t) +  e_{max}(t)}{2}.
\end{equation}

To define these envelopes, Huang et al (1998) used a cubic interpolation
between the local extrema of the signal. This is illustrated in Fig.
\ref{fig:TwoModes} by the dashed lines, whilst the local mean value
is represented by the solid line. Once the trend $r(t)$ has been evaluated,
it is then easy to extract the detail $d(t)$ via Eq.
(\ref{eq:EMDbasics}).

The comparisons of the trend $r(t)$ and the detail $d(t)$, estimated from the
signal $s(t)$ by means of the EMD algorithm, with the low- and
high-frequency components (see Eq. (\ref{eq:ExSignal})) are given in Figs.
\ref{fig:TwoModesTrend} and \ref{fig:TwoModesDetail}, respectively. The excellent
agreement shown in these plots evidences the potentiality of the EMD.

\subsection{The generalization}

This procedure can be extended to any signal $s(t)$ containing more than
two characteristic frequencies. In that case, the original signal $s(t)$
is first decomposed into two contributions according to Eq. (\ref{eq:EMDbasics}).
Then, the trend $r(t)$ resulting from this decomposition is used as a "new"
signal onto which the EMD method is applied in turn. This procedure is
reiterated until no more frequency can be identified. The original
signal can therefore be expressed as follows

\begin{equation}
s(t) = \sum^{N}_{k = 1} \mbox{IMF}_{k}(t) + r_N(t),
\end{equation}

where $N$ is the total number of iterations, $r_N(t)$ is the final residue and
$\mbox{IMF}_k(t)$ is the $k^{\mbox{th}}$ Intrinsic Mode Function, i.e.
the trend of the $(k-1)^{\mbox{th}}$ iteration, extracted via the procedure described
above. Given that the EMD is by definition a data-driven technique, therefore
locally self-adaptive, two important remarks have to be made to emphasize the
main differences between EMD and standard approaches:

\begin{enumerate}
\item[i)] the shape of the IMFs is not assumed \textit{a priori} unlike
the Fourier transform.
\item[ii)] the extraction of the IMFs is not conditioned by a statistical
convergence unlike the Proper Orthogonal Decomposition for instance.
\end{enumerate}

These two degrees of freedom provide to the EMD method an attractive potential
to study forced turbulence in the sense that the IMFs act as non-linear
dynamic filters. This property is straightly related to the concept of
instantaneous frequency. In order to define properly the instantaneous frequency,
the IMFs have to fulfill two conditions (see \cite{Huangetal1998}):
  
\begin{enumerate}
\item[i)] the total number of zero crossings and extremes differ by one at most.
\item[ii)] the local mean value defined as the average between the lower and the
upper envelopes is zero.
\end{enumerate}

Huang et al. \cite{Huangetal1998} developed a sifting procedure consisting in repeating,
for each iteration, the EMD analysis until the IMF complies with both enumerated
conditions. In the present study, we have restricted, for practical reason,
the number of sifting repetition to 15 but no significant change has been observed
when increasing this value.

\section{The methodology}
\label{sec:method}

As mentioned earlier, the performances of the EMD algorithm are evaluated on
an experimental turbulent signal which is artificially perturbed by a
numerical long-period flapping. This section describes the strategy we have
adopted to assess the relevance of the EMD to separate the perturbation
from the reference signal.

\begin{figure}[htbp]
\centering
\subfigure[]
{\includegraphics[width=0.45\textwidth]{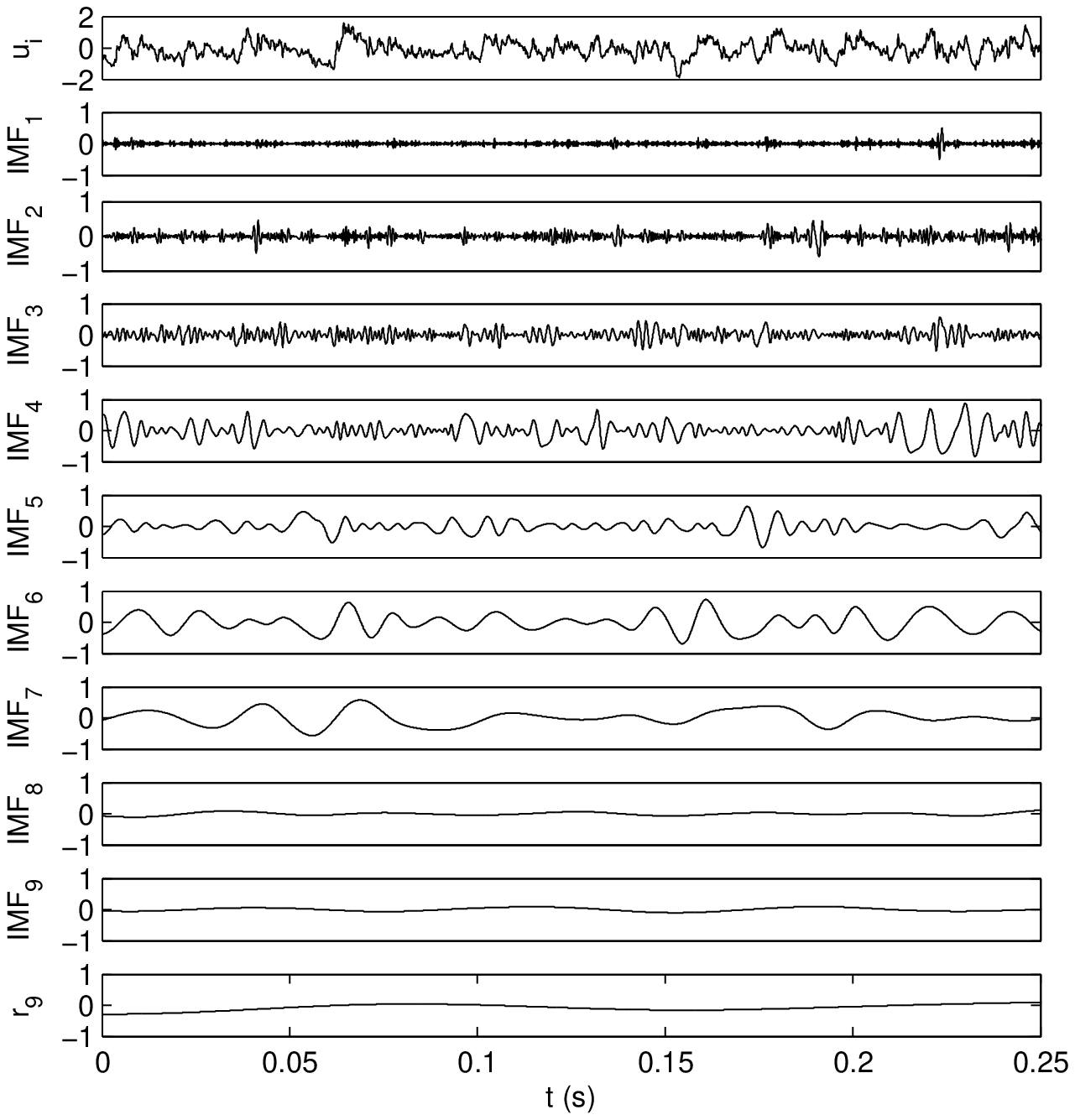}
\label{fig:EMDui}}
\hspace{0.2cm}
\subfigure[]
{\includegraphics[width=0.45\textwidth]{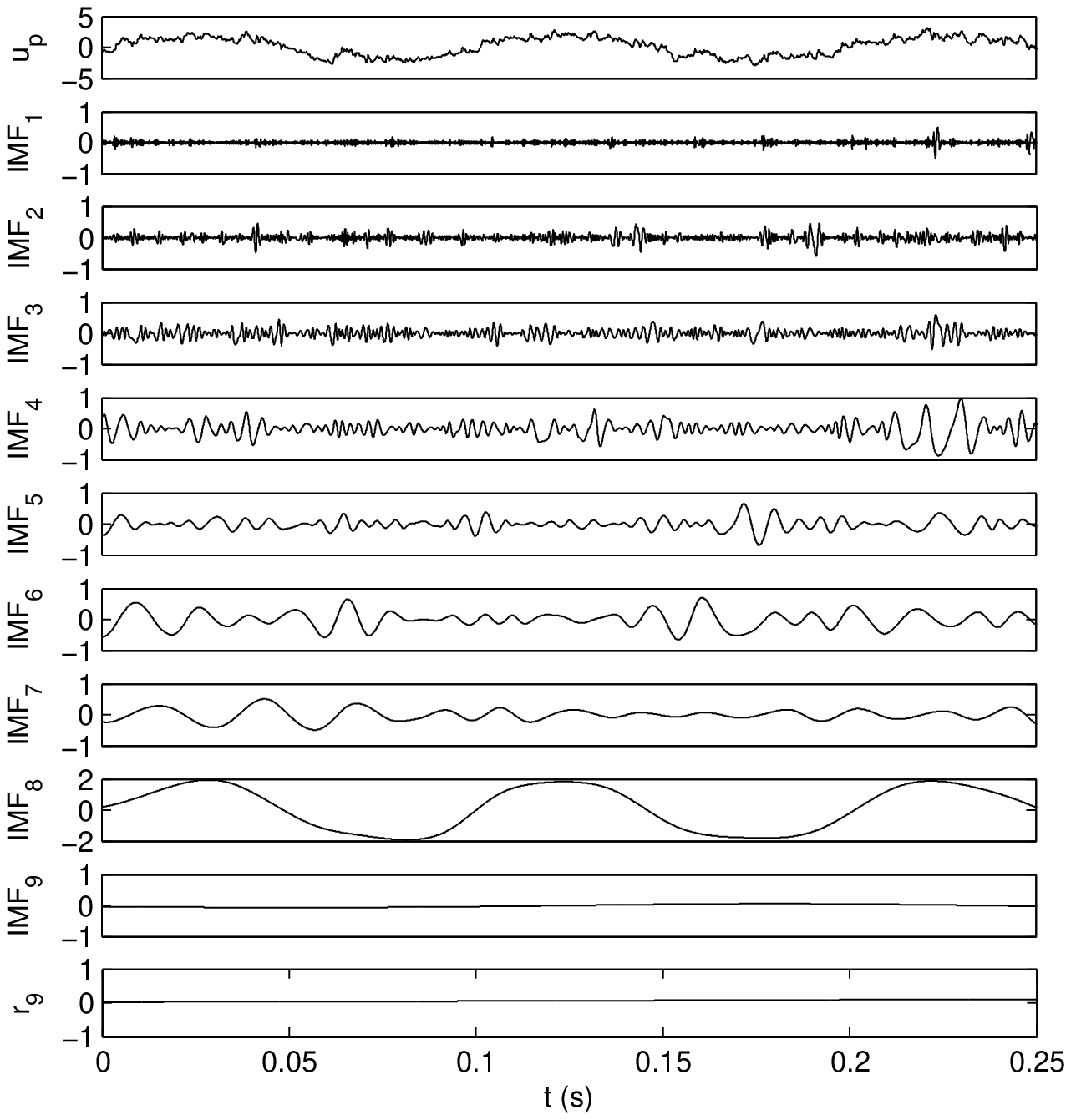}
\label{fig:EMDup10Hz}}
\caption{EMD analysis of the reference signal $u_i(t)$ (a)
and of a perturbed signal (b) with $f_p = 10$Hz and
$a_p = 2$ m/s. In both cases, the top plot represents
the analyzed signal.}
\end{figure}

\subsection{The reference signal}

The reference signal, used in this study, is a velocity fluctuation time-series
measured in a fully-developed decaying turbulence. The experiment was
carried out in an open-loop vertical wind tunnel of the laboratory CORIA
(Rouen). The turbulence generator is a perforated plate with an uniform
mesh size $M \approx 25mm$ set at the inlet of the working section. A detailed
investigation of the turbulent flow produced downstream the perforated
plate has been reported elsewhere by Mazellier et al. \cite{Mazellieretal2010}. The streamwise
velocity fluctuation sample $u_i(t)$ used here has been measured via
Laser Doppler Velocimetry at approximately 15$M$ downstream the perforated
plate where the flow is nearly homogeneous and isotropic
(see \cite{Mazellieretal2010} for more details). The main physical properties
of the turbulent flow at this location are summarized in Table \ref{tab:flow}.

\begin{table}[htbp]
\caption{Main physical characteristics of the reference turbulent flow}
\label{tab:flow} 
\begin{tabular}{cccccc}
\hline\noalign{\smallskip}
$U^a$ (m/s) & $\left<u_i^2\right>^{1/2}$ (m/s) & $L_i^b$ (mm) & $\lambda_i^c$ (mm) & $\eta_i^d$ & $Re_\lambda^e$  \\
\noalign{\smallskip}\hline\noalign{\smallskip}
4.1 & 0.57 & 9.4 & 1.6 & 0.1 & 59 \\
\noalign{\smallskip}\hline
\end{tabular}\\
$^a$ Streamwise mean velocity,\\
$^b$ Integral length-scale computed from Eq. (\ref{eq:L}),\\
$^c$ Taylor microscale computed from Eq. (\ref{eq:lambda}),\\
$^d$ Kolmogorov scale computed from Eq. (\ref{eq:eta}),\\
$^e$ Taylor-based Reynolds number ($\equiv \left<u_i^2\right>^{1/2} \lambda_i / \nu$ with
$\nu$ the kinematic viscosity).
\end{table}

Note that the blockage ratio $\sigma$ of the perforated plate (defined as the ratio between the
blocked area and the wind-tunnel's section) used by Mazellier et al. \cite{Mazellieretal2010}
is significantly higher ($\sigma =0.67$) than that usually reported for bi-plane regular grids
(see e.g. \cite{ComteBellotCorrsin1966} with $\sigma = 0.34$). This difference
yields to a much higher turbulence intensity $\left<u_i^2\right>^{1/2}/U \approx 13$\%
(where $\left<\mbox{ }\right>$ denotes a time-averaging and $U$ the mean velocity) than that of
standard grid-generated turbulence. However, regarding the dimension of the experiment,
the turbulent flow is restricted to moderate
Taylor-based Reynolds number $Re_\lambda$ ($\equiv \left<u_i^2\right>^{1/2} \lambda_i / \nu$ with
$\nu$ the kinematic viscosity).

\subsection{The perturbation}

In this study, the general form of the numerical perturbation $p_i(t)$
is the following

\begin{equation}
p_i(t) = a_p(t) \sin \left(\phi(t)\right),
\end{equation}

where $a_p$ and $\phi$ are the amplitude and the phase of the perturbation,
respectively. The perturbation frequency $f_p$ and the phase are related
by

\begin{equation}
\frac{d\phi}{dt} = 2 \pi f_p(t).
\end{equation} 

In the following, we report two cases mimicking either a "mono-component"
or a "multi-component" flapping. The "mono-component" flapping is characterized by
a constant perturbation frequency (sine wave), whilst $f_p$ is time-dependent,
i.e. $f_p = f_p(t)$, for the case of the "multi-component" flapping (linear
chirp). 

The perturbed signal $u_p(t)$ under analysis is therefore

\begin{equation}
u_p(t) = u_i(t) + p_i(t). 
\end{equation}

The goal of the present
work is to assess the ability of the EMD method to extract
a long-period flapping from the turbulent signal. For that reason,
the frequency range of the perturbation is much lower than the typical
frequency $\left<u_i^2\right>^{1/2}/L_i$ of the integral length-scale
($\approx 60$Hz).

A typical EMD analysis of both the reference signal and a perturbed signal
is displayed in Figs. \ref{fig:EMDui} and \ref{fig:EMDup10Hz},
respectively. For that specific example, the amplitude $a_p$ and the frequency
$f_p$ of the perturbation are equal to 2 m/s and 10 Hz, respectively. The
presence of the perturbation is evident on the top plot of Fig.
\ref{fig:EMDup10Hz} where a long-period variation superimposes to the
dynamics of the reference signal (see top plot of Fig. \ref{fig:EMDui}).

These plots suggest that the characteristic frequencies
of the IMFs decrease with increasing IMF's number meaning that the low frequency
dynamics is mainly concentrated in the higher order IMFs. This
statement is well supported by a close comparison of the IMFs time-series
computed from both the reference and the perturbed signals. For that particular
example, one can see that the IMFs are almost identical until $\mbox{IMF}_6$.
For higher order IMFs, significant discrepancies can be observed, especially for
$\mbox{IMF}_8$. A careful examination of that specific IMF with the perturbed signal
(top plot in Fig. \ref{fig:EMDup10Hz}) reveals a strong resemblance in shape at
low frequency.

\subsection{The "resemblance" criterion}

\begin{figure}[htbp]
\centering
\subfigure[]
{\includegraphics[width=0.65\textwidth]{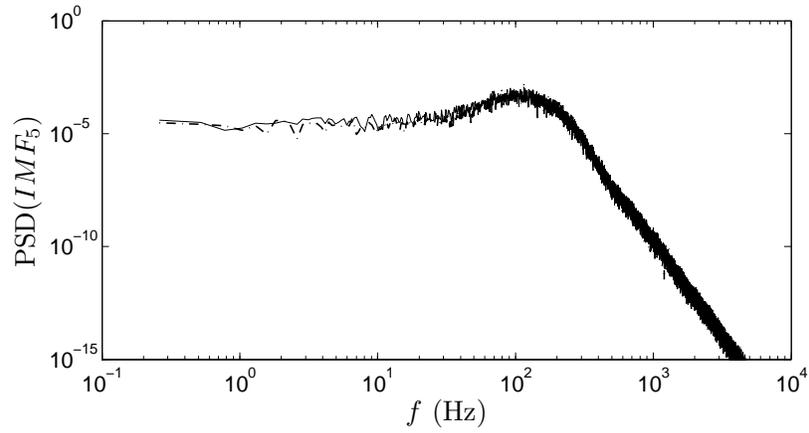}
\label{fig:SpecIMF5f10a2}}
\vspace{0.1cm}
\subfigure[]
{\includegraphics[width=0.65\textwidth]{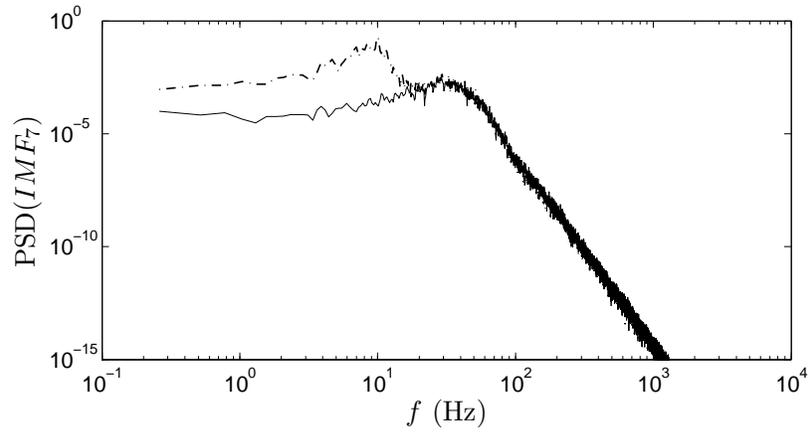}
\label{fig:SpecIMF7f10a2}}
\vspace{0.1cm}
\subfigure[]
{\includegraphics[width=0.65\textwidth]{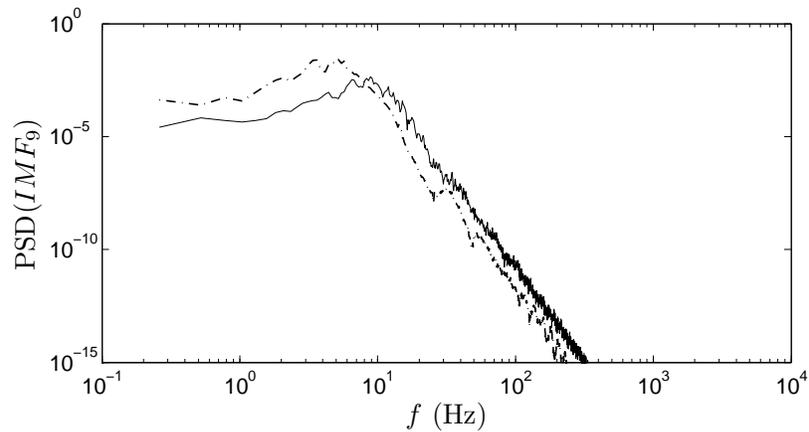}
\label{fig:SpecIMF9f10a2}}
\caption{1D energy spectra of the (a) 5$^{\mbox{th}}$ IMF,
(b) 7$^{\mbox{th}}$ IMF and (c) 9$^{\mbox{th}}$ IMF time-series extracted
from the reference signal (solid lines) and the perturbed
signal (broken lines with dots) with $f_p = 10$Hz and $a_p = 2$m/s.}
\end{figure}

The simple example shown hereinbefore indicates that only a limited number of
IMFs are polluted by the numerical perturbation. The issue arising from that
observation relies on the relevant discrimination between "good", i.e. unpolluted,
and "bad", i.e. polluted, IMFs. Foucher and Ravier \cite{FoucherRavier2010} pointed out that this
discrimination requires to define a criterion being robust and unbiased as well.
In the case of the sum of two tones, Rilling and Flandrin \cite{RillingFlandrin2008} introduced
a performance measurement based on the relative error of the reconstructed
signal compared to the original one. Although this method shown a great
efficiency, it requires to know \textit{a priori} the reference signal
which, obviously, is not available when studying real turbulent signals.
Foucher and Ravier \cite{FoucherRavier2010} recovered successfully the main large scale
features of a turbulent flow exited by an external forcing by defining
a tunable cut-off frequency. Their results revealed a strong dependence 
on that cut-off frequency enabling to determine an efficient value. Even
though their approach gave good results, the assumption
that the cut-off frequency is unique for each IMF is questionable. This is
evidenced in Figs. \ref{fig:SpecIMF5f10a2}-\ref{fig:SpecIMF9f10a2} which
show the Power Spectral Densities (PSD) of three IMFs extracted from both
the reference signal and the perturbed signal exampled hereinbefore (see
Figs. \ref{fig:EMDui} and \ref{fig:EMDup10Hz}). As expected, the low-order IMF ($5^{\mbox{th}}$ IMF, here)
is unaffected by the perturbation (see Fig. \ref{fig:SpecIMF5f10a2}). One can
see that the effect of the perturbation increases when increasing the IMF's
order. While the definition of a cut-off frequency may be appropriate in the
case illustrated in Fig. \ref{fig:SpecIMF7f10a2} where a strong peak of energy
appears at the excitation frequency, this strategy is irrelevant for the
highest-order IMF shown in Fig. \ref{fig:SpecIMF9f10a2} due to the frequency
shift accounted for the non-linear feature of the EMD algorithm.

\begin{figure*}[htbp]
\centering
\subfigure[]
{\includegraphics[width=0.55\textwidth]{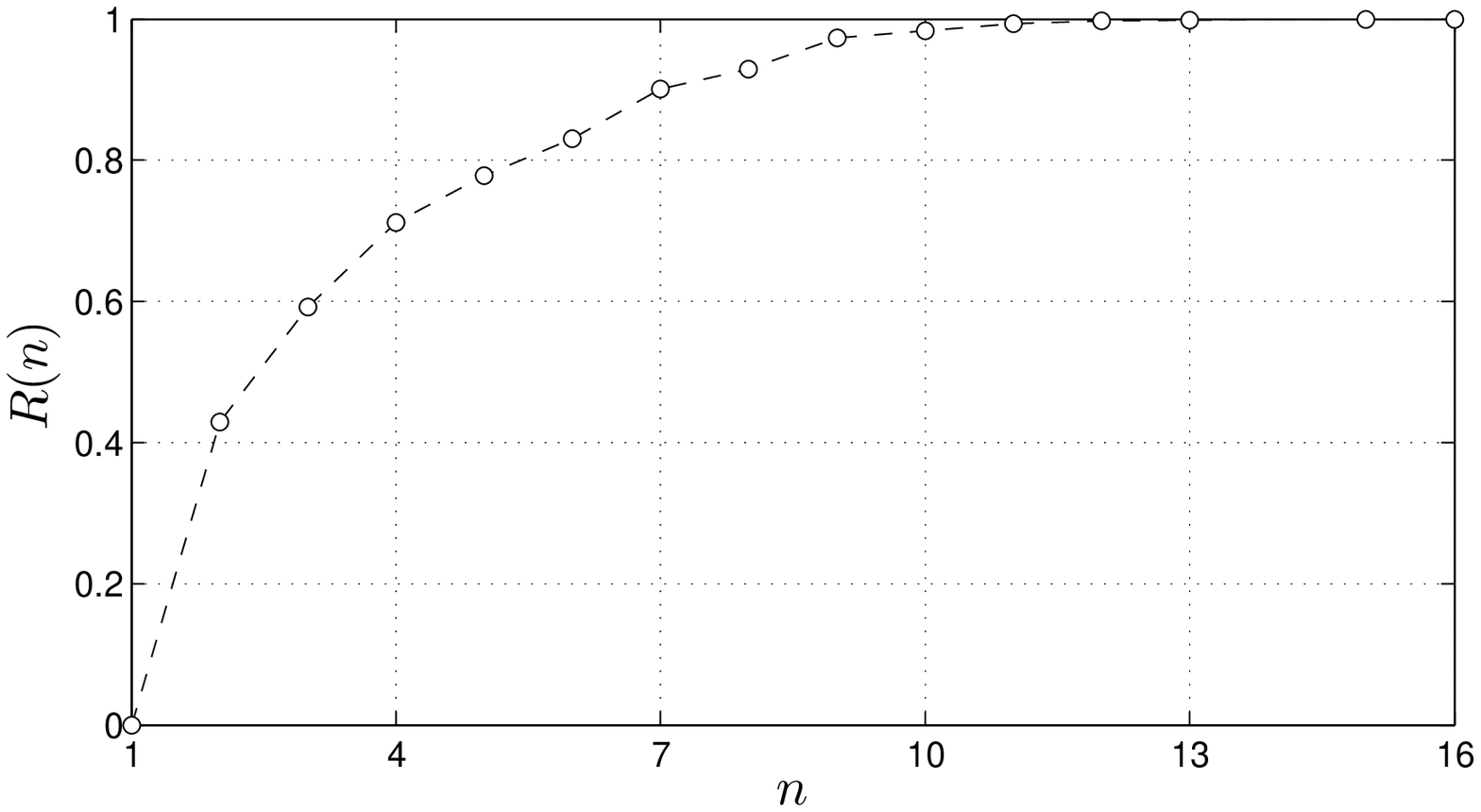}
\label{fig:Rf0}}
\hspace{0.2cm}
\subfigure[]
{\includegraphics[width=0.55\textwidth]{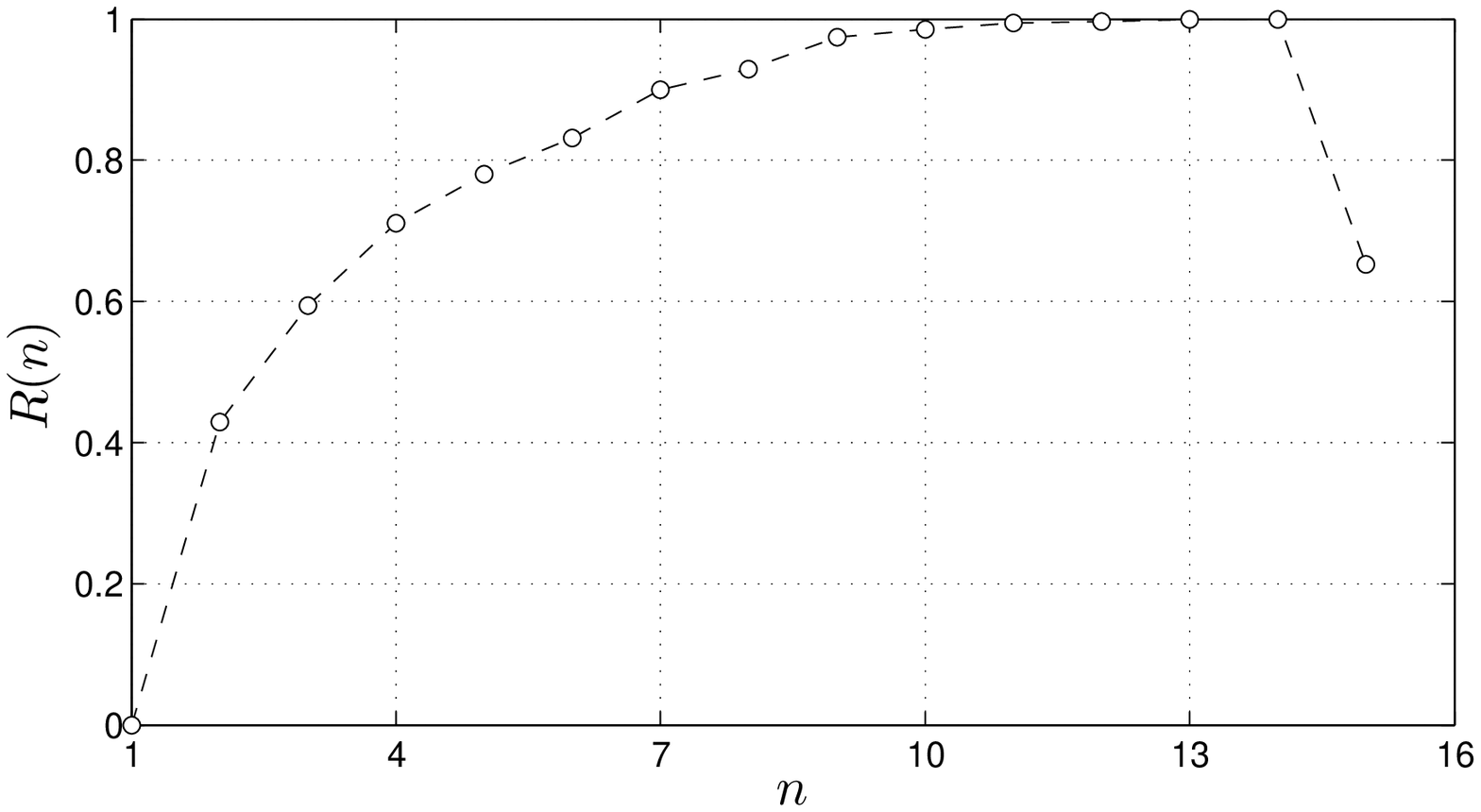}
\label{fig:Rf002}}
\vspace{0.1cm}
\subfigure[]
{\includegraphics[width=0.55\textwidth]{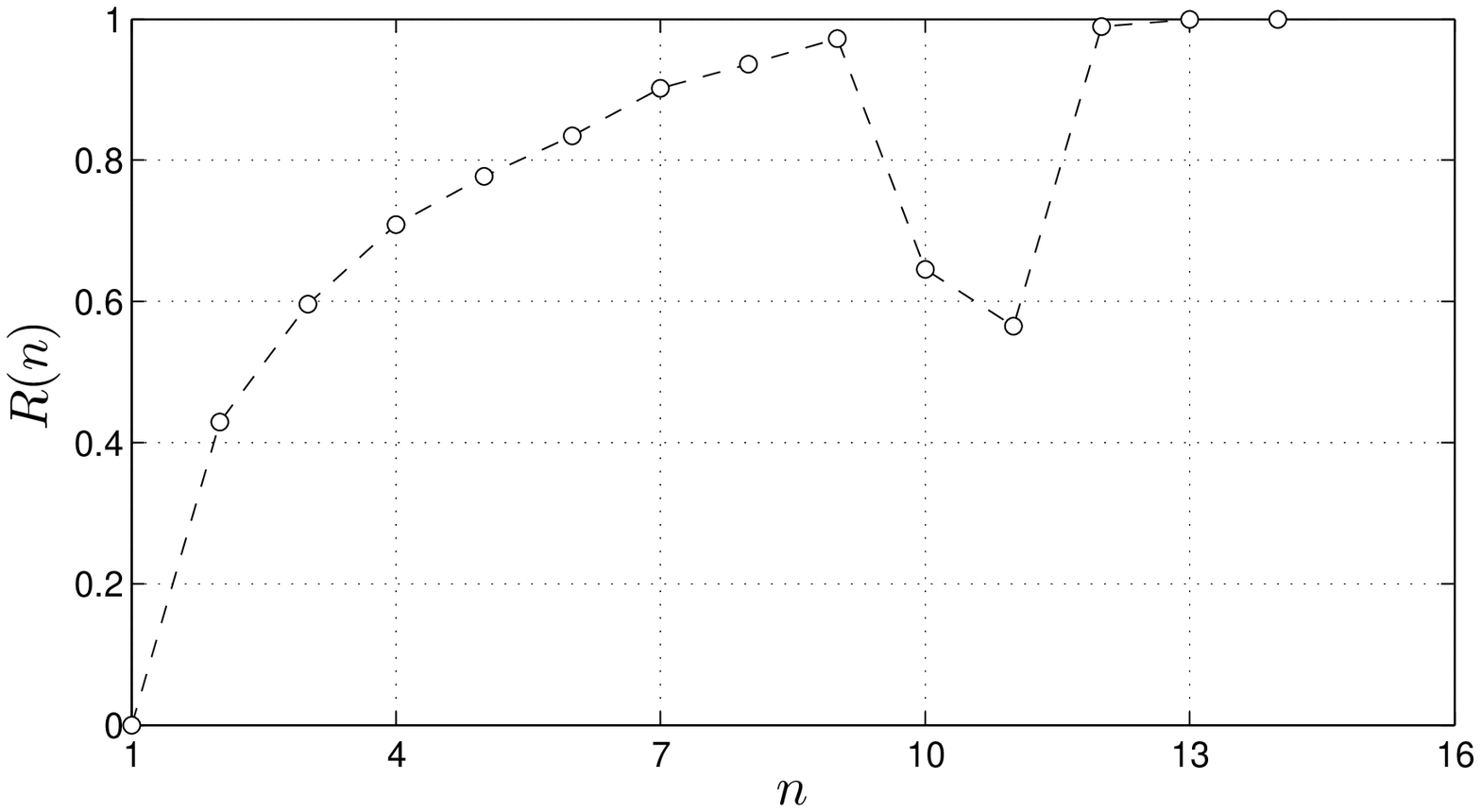}
\label{fig:Rf1}}
\hspace{0.2cm}
\subfigure[]
{\includegraphics[width=0.55\textwidth]{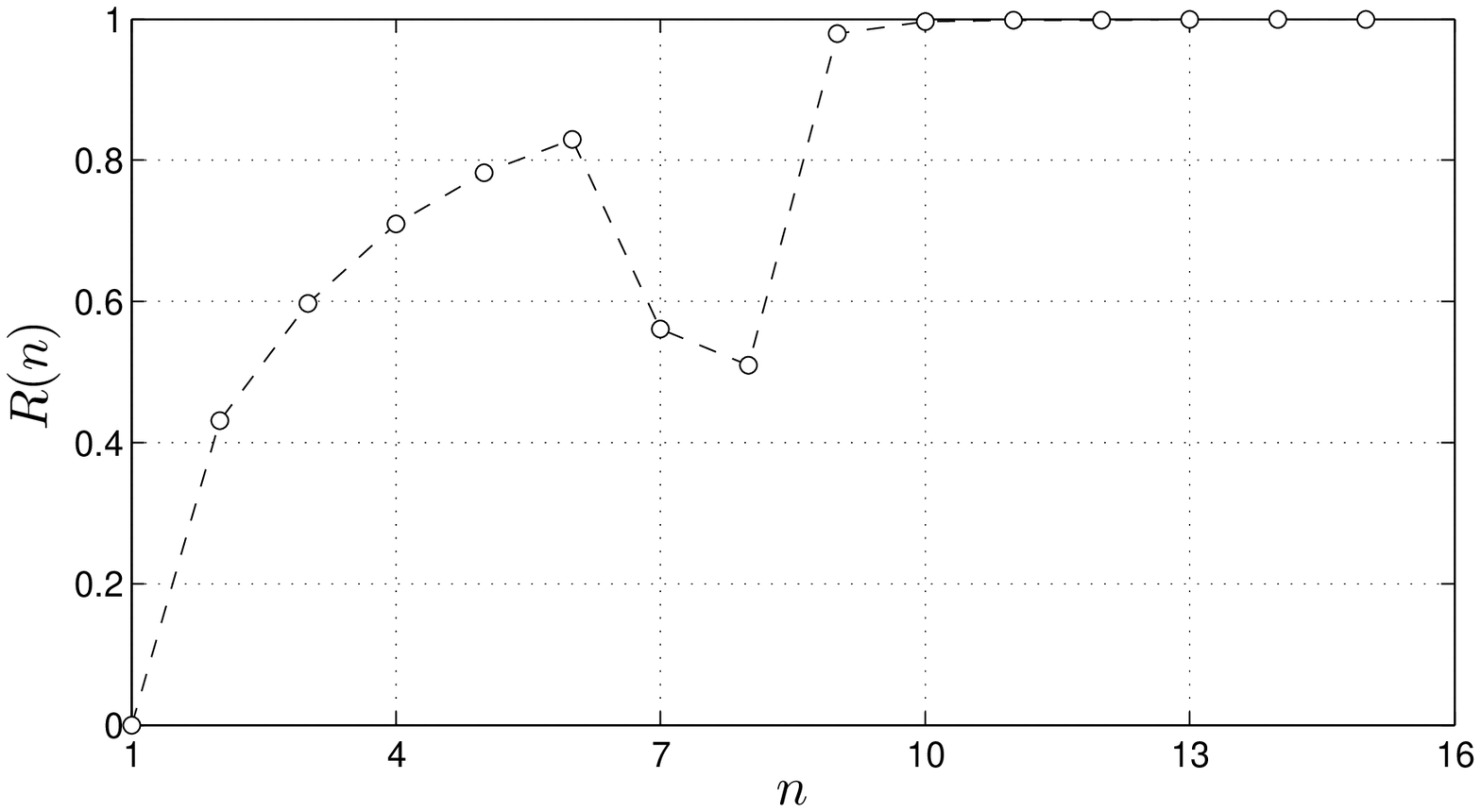}
\label{fig:Rf10}}
\caption{Evolution of the "resemblance" criterion $R(n)$
as a function of the reconstruction number $n$ for (a) the reference
signal and perturbed signals: (b) $f_p = 0.02$Hz, (c) $f_p = 1$Hz
and (d) $f_p = 10$Hz. For each perturbed signal, the perturbation
amplitude is constant ($a_p = 2$m/s).}
\end{figure*}

To address the issue of the separation between polluted and
unpolluted IMFs, we introduce a new criterion, referred to as
the "resemblance" criterion, $R(n)$ such that

\begin{equation}
\left\{
\begin{array}{ll}
R\left(n\right) = 0 & \mbox{  if } n = 1, \\
&\\
R\left(n\right) = \frac{\left\langle u_n(t) u_{n-1}(t)\right\rangle}
{\sqrt{\left\langle u_n(t)^2\right\rangle}
\sqrt{\left\langle u_{n-1}(t)^2\right\rangle}}
 & \mbox{  if } 1 < n \leq N,
\end{array}
\right.
\label{eq:R}
\end{equation}

where the reconstructed signals $u_n(t)$ are defined as follows

\begin{equation}
\left\{
\begin{array}{ll}
u_n(t) = \sum^{n}_{k=1} \mbox{IMF}_k(t) & \mbox{  if } n < N, \\
&\\
u_n(t) = \sum^{n}_{k=1} \mbox{IMF}_k(t) + r_N & \mbox{  if } n = N,
\end{array}
\right.
\label{eq:Recon}
\end{equation}

with $n$ the reconstruction number. Eq. (\ref{eq:Recon}) represents the
step-by-step reconstruction of the signal $u_p(t)$ by adding more
and more IMFs. Therefore, the "resemblance" criterion $R(n)$ evaluates
the statistical dependence between the reconstructed signals resulting from
two successive steps. Fig. \ref{fig:Rf0} shows the evolution of $R\left(n\right)$
in the specific case of the reference signal (i.e. unperturbed).
Starting from 0 by definition, $R\left(n\right)$ increases monotonically with
respect to $n$ until it saturates to 1 when $n$ tends towards $N$.

The efficiency of the "resemblance" criterion to discriminate between
the perturbation and the reference signal relies on the fact that their
dynamics (or more generally their physical properties) differs noticeably,
in other words, their statistical dependence has to be small enough. This
statement is emphasized in Figs. \ref{fig:Rf002}-\ref{fig:Rf10} which
display the typical evolution of $R\left(n\right)$ for several perturbed
signals. For these examples, the perturbation frequency $f_p$
is respectively $0.02$Hz, $1$Hz and $10$Hz, whilst
the perturbation amplitude is kept constant ($a_p = 2$ m/s). For each case,
even though the global trend of $R(n)$ is similar to that obtained
for the reference signal, a significant drop of $R\left(n\right)$ is
easily detected. This fall results from the addition of a polluted
IMF (i.e. a different dynamics) to the reconstructed signal which
leads, therefore, to a decrease of the cross-correlation, i.e.
the statistical dependence, between the $n^{\mbox{th}}$ and the $(n-1)^{\mbox{th}}$
reconstruction steps. Then, the "resemblance" criterion increases again
reflecting the fact that the reconstructed signal becomes more and more
resemblant to the perturbed signal. Note that for the specific case
of $f_p = 10$Hz, the drop of $R(n)$ coincides with the 7$^{\mbox{th}}$ IMF in
agreement with the spectral analysis displayed in Fig. \ref{fig:SpecIMF7f10a2}.

\subsection{The rejection procedure}

Figs. \ref{fig:Rf002}-\ref{fig:Rf10} show that the reconstruction number $n$
where the fall of $R(n)$ occurs depends on $f_p$. This easily
identifiable position is used to define a cut-off mode $k_c$ beyond
which the IMFs are systematically discarded. It is important to remark
that this procedure significantly differs from that used by Foucher and Ravier \cite{FoucherRavier2010}.
Indeed, the IMFs for which $k \geq k_c$ are all affected by the perturbation
(see e.g. Fig. \ref{fig:SpecIMF9f10a2}) and, therefore, are not used
in the recovering procedure. This point is further discussed in Sec. 
\ref{sec:results}.

Finally, we define the recovered signal $u_r(t)$ such as

\begin{equation}
u_r(t) = \sum^{k_c}_{k=1} \mbox{IMF}_k(t).
\label{eq:ur}
\end{equation}

This signal can then be used to estimate, \textit{a posteriori}, the
perturbation

\begin{equation}
p_r(t) = u_p(t) - u_r(t),
\label{eq:pr}
\end{equation}

with $p_r(t)$ the estimated perturbation. The recovered signal $u_r(t)$ can be
assimilated to a "high-pass filtered" version of $u_p(t)$. Here, the term "high-pass
filtering" has to be understood in a different way than usual. Indeed, as evidenced
in Figs. \ref{fig:SpecIMF5f10a2}-\ref{fig:SpecIMF9f10a2}, each IMF cover a broad range
of frequency in the spectral space. In particular, a significant part of low-frequency
energy (that of the turbulent large-scales) is contained in unpolluted IMFs. Consequently,
unlike usual high-pass filtering, the recovering procedure does not annihilate totally
the energy below a given cut-off frequency. This is illustrated in Sec. \ref{sec:results}
where the performances of the recovering procedure is evaluated in various conditions.

\section{Results}
\label{sec:results}

This section presents a qualitative and quantitative investigation
assessing the EMD performances in discriminating the velocity signal
from the perturbation. First, a parametric study, in the case of
a "mono-component" perturbation (sine wave), is reported. Then, the
influence of the perturbation frequency is analyzed, in details, by
investigating the scale-by-scale turbulent energy by means of
energy spectra and structure functions. A particular attention is
given to the main turbulence properties at large- and small-scales.
Finally, the results of the extension to a time-dependent perturbation
(linear chirp) are provided.

\subsection{The "mono-component" flapping}

\begin{figure}[htbp]
\centering
\includegraphics[width=0.48\textwidth]{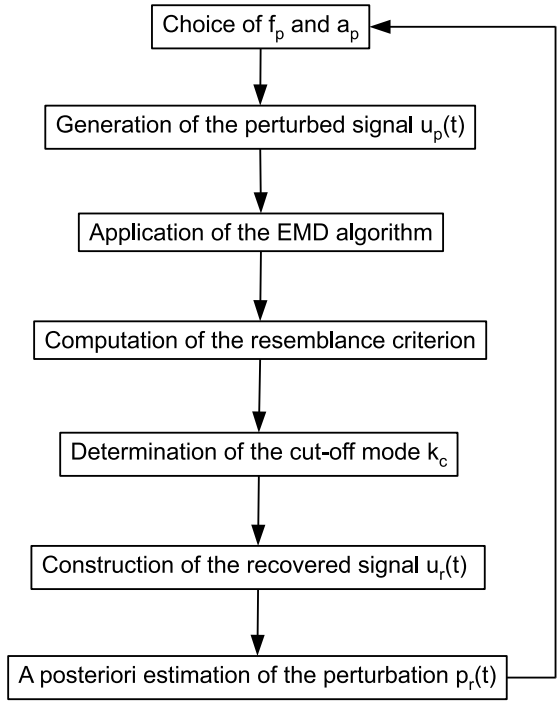}
\caption{Schematic of the procedure followed to recover
$u_r(t)$ and $p_r(t)$ from the perturbed signal $u_p(t)$.}
\label{fig:Procedure}
\end{figure}

An extensive investigation of the influence of both parameters
$a_p$ and $f_p$ has been conducted following the procedure described
in Fig. \ref{fig:Procedure} in the case of a "mono-component"
perturbation. At each iteration of this parametric study, a couple
($a_p$, $f_p$) is fixed and the recovering algorithm described in Sec.
\ref{sec:method} is applied on the perturbed signal in order to extract,
\textit{a posteriori}, the velocity signal and the perturbation
according to Eqs. (\ref{eq:ur}) and (\ref{eq:pr}). In this study,
the ratio of the perturbation energy ($a_p^2/2$)
to the turbulent kinetic energy ($\left<u_i^2\right>$) is lower than 25.
Furthermore, in order to simulate a long-period flapping,
the maximum perturbation frequency $f_p$ is about 6 times smaller
than the frequency of the energy-contained eddies ($\equiv
\left<u_i^2\right>^{1/2}/L_i$).

The quality of the recovering algorithm can be assessed by means of the
coefficient $C$ defined as follows

\begin{equation}
C = \frac{\left\langle u_i(t) u_r(t)\right\rangle}
{\sqrt{\left\langle u_i^2\right\rangle}
\sqrt{\left\langle u_r^2\right\rangle}}.
\label{eq:C}
\end{equation}

\begin{figure*}[htbp]
\centering
\subfigure[]
{\includegraphics[width=0.46\textwidth]{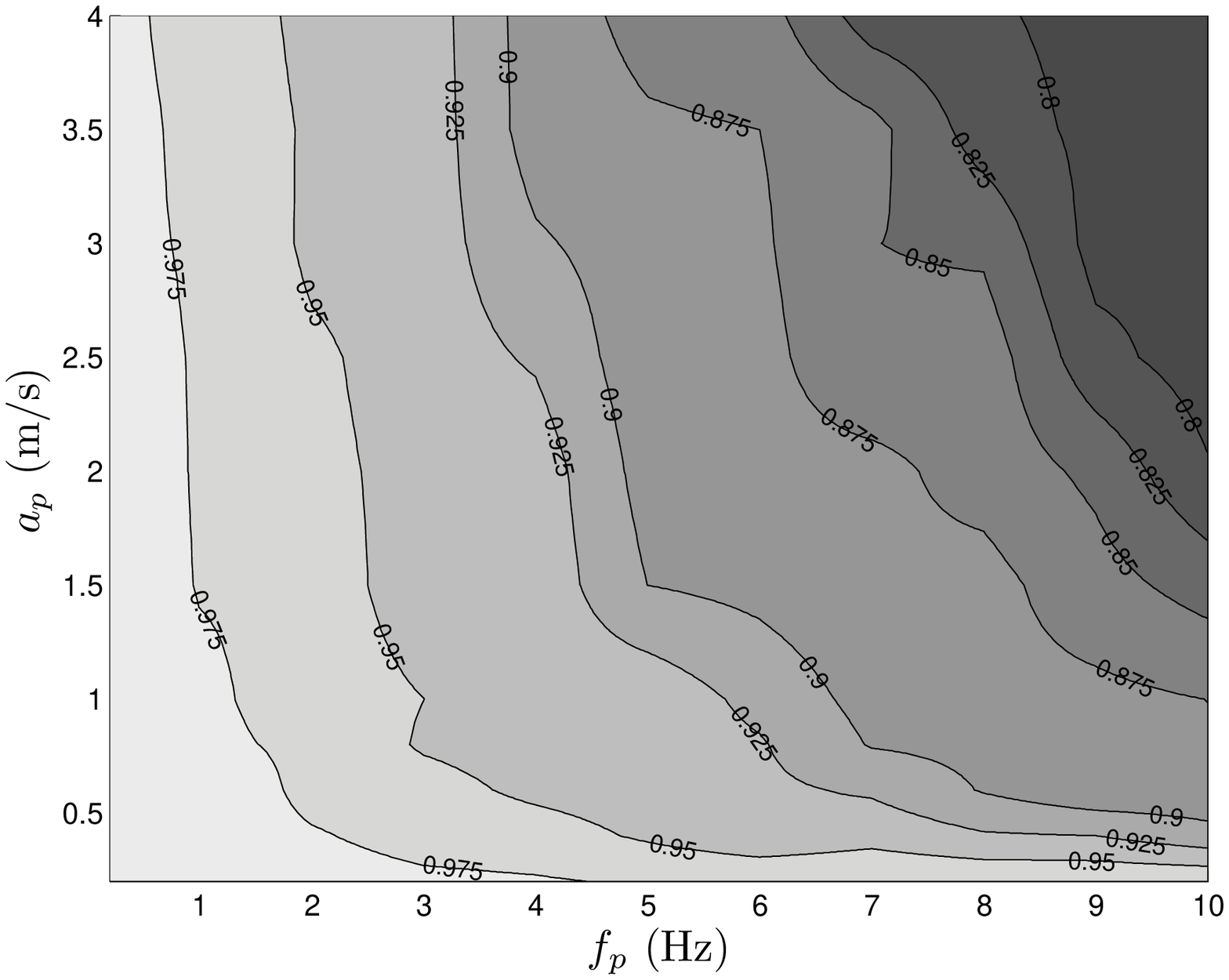}
\label{fig:C_2e15}}
\hspace{0.1cm}
\subfigure[]
{\includegraphics[width=0.46\textwidth]{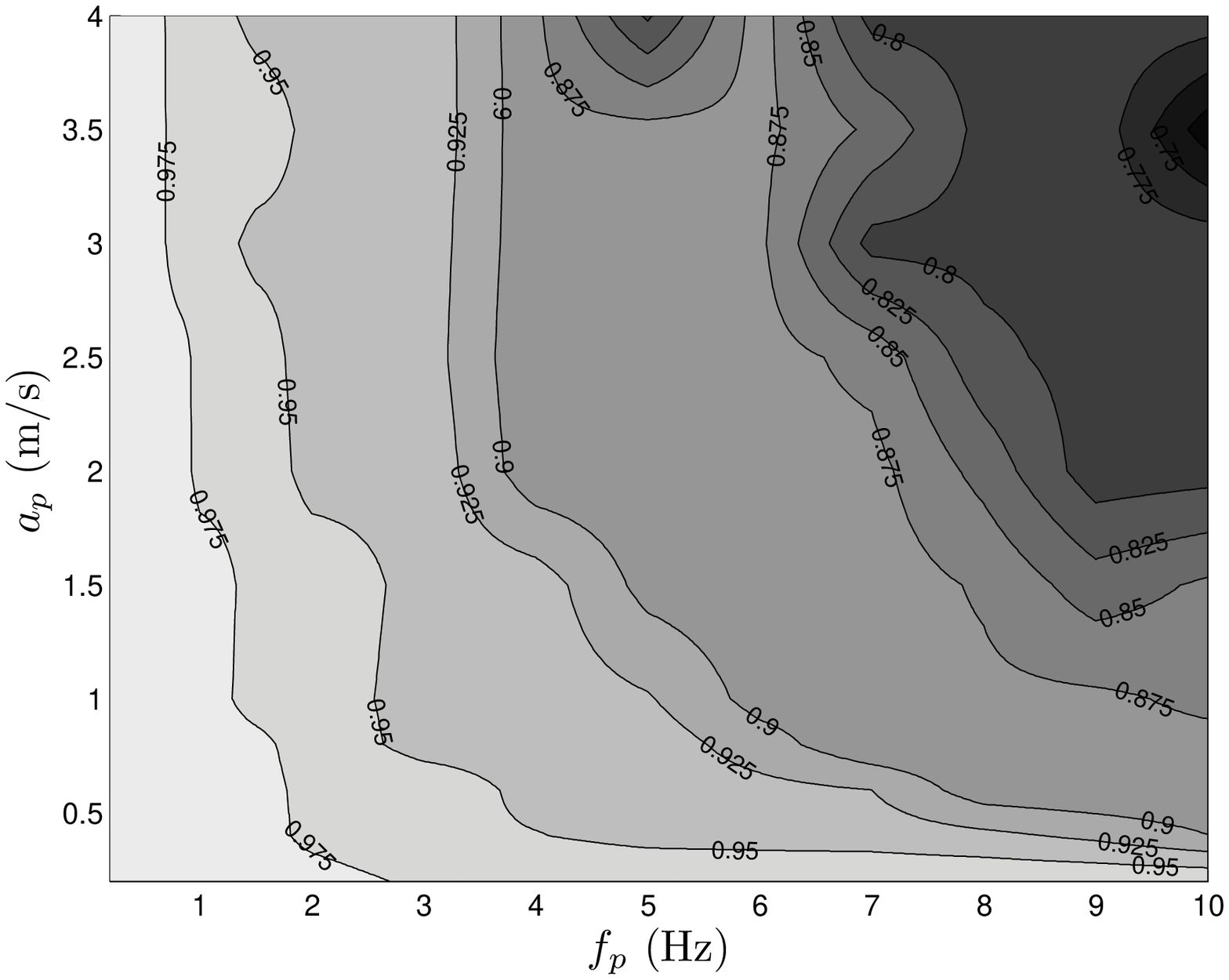}
\label{fig:C_2e18}}
\caption{Contour plot of the recovering quality indicator $C$ with respect
to the parameters of the "mono-component" perturbation. Length of the
reference signal: (a) $t_m \approx$ 2s ($2^{15}$ samples), (b) $t_m \approx$ 15s
($2^{18}$ samples).}
\end{figure*}

The variation of $C$ as a function of the couple ($a_p, f_p$) is plotted
in Figs. \ref{fig:C_2e15} and \ref{fig:C_2e18} for a signal duration $t_m$
of about 2s ($2^{15}$ samples) and 15s ($2^{18}$ samples),
respectively.

The high values of $C$ (mostly $\geq 0.8$) observed in both plots reflect
that the recovering procedure based on the EMD algorithm performs remarkably 
well over the whole range of perturbation amplitude and frequency tested.
Excluding the area where $a_p \geq 3$, a close comparison
between both plots indicates that the influence of the signal length on the
recovering quality is weak. Therefore, in the following, only results
obtained for $t_m \approx$ 15s ($2^{18}$ samples) are reported.

Overall, the recovering procedure seems less sensitive to $a_p$
than to $f_p$ and it turns out that the quality
of the recovering procedure decreases with increasing $f_p$.
In order to analyze this feature, we now focus the discussion,
on three typical "mono-component" perturbations differing only by their
frequency, i.e. $f_p = 0.02$Hz, 1Hz and 10Hz respectively, while their
amplitude is kept constant ($a_p = 2$m/s). These specific frequencies cover
the entire range of flapping frequency simulated in this study.

\begin{figure*}[htbp]
\begin{minipage}[c]{.55\linewidth}
\subfigure[]
{\includegraphics[width=0.9\textwidth]{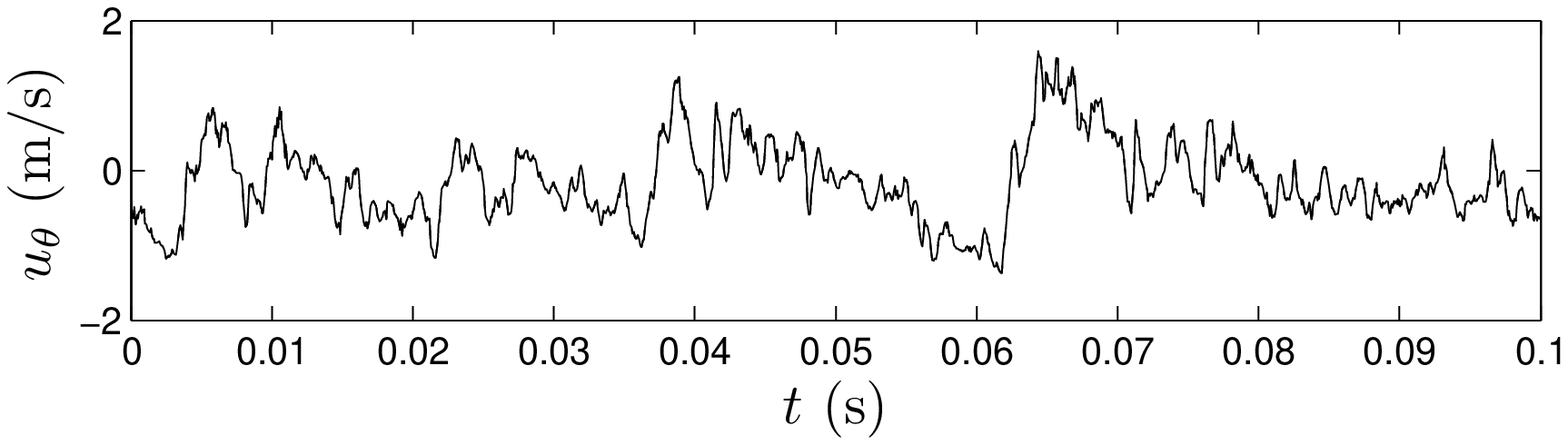}
\label{fig:uf002a2}} 
\subfigure[]
{\includegraphics[width=0.9\textwidth]{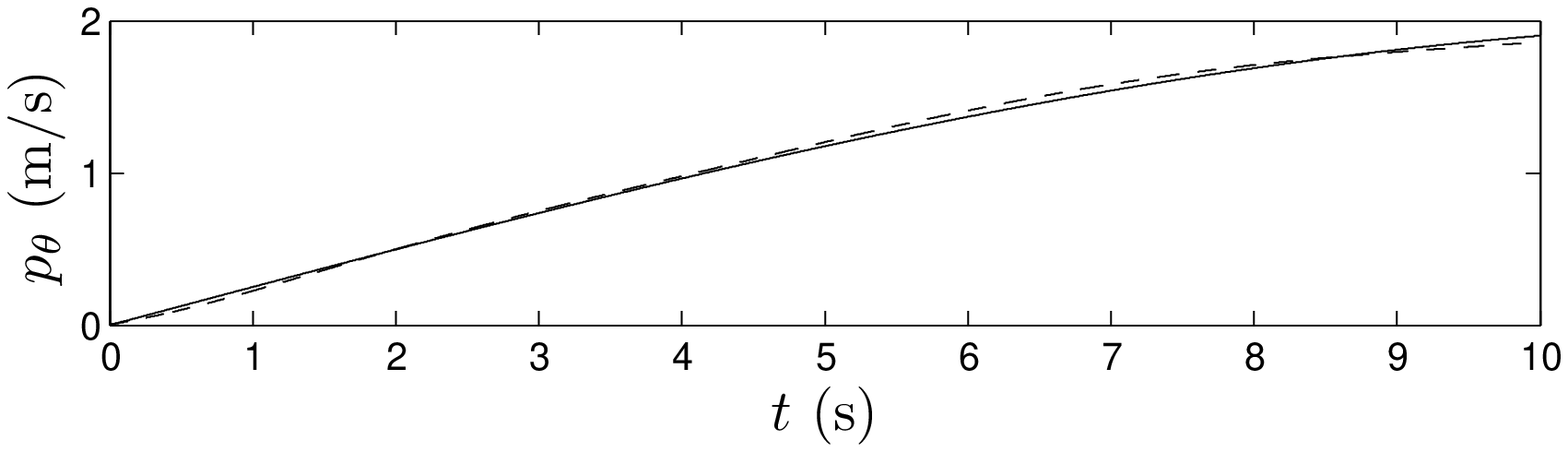}
\label{fig:pf002a2}}
\end{minipage}
\hfill
\begin{minipage}[c]{.41\linewidth}
\subfigure[]
{\includegraphics[width=1\textwidth]{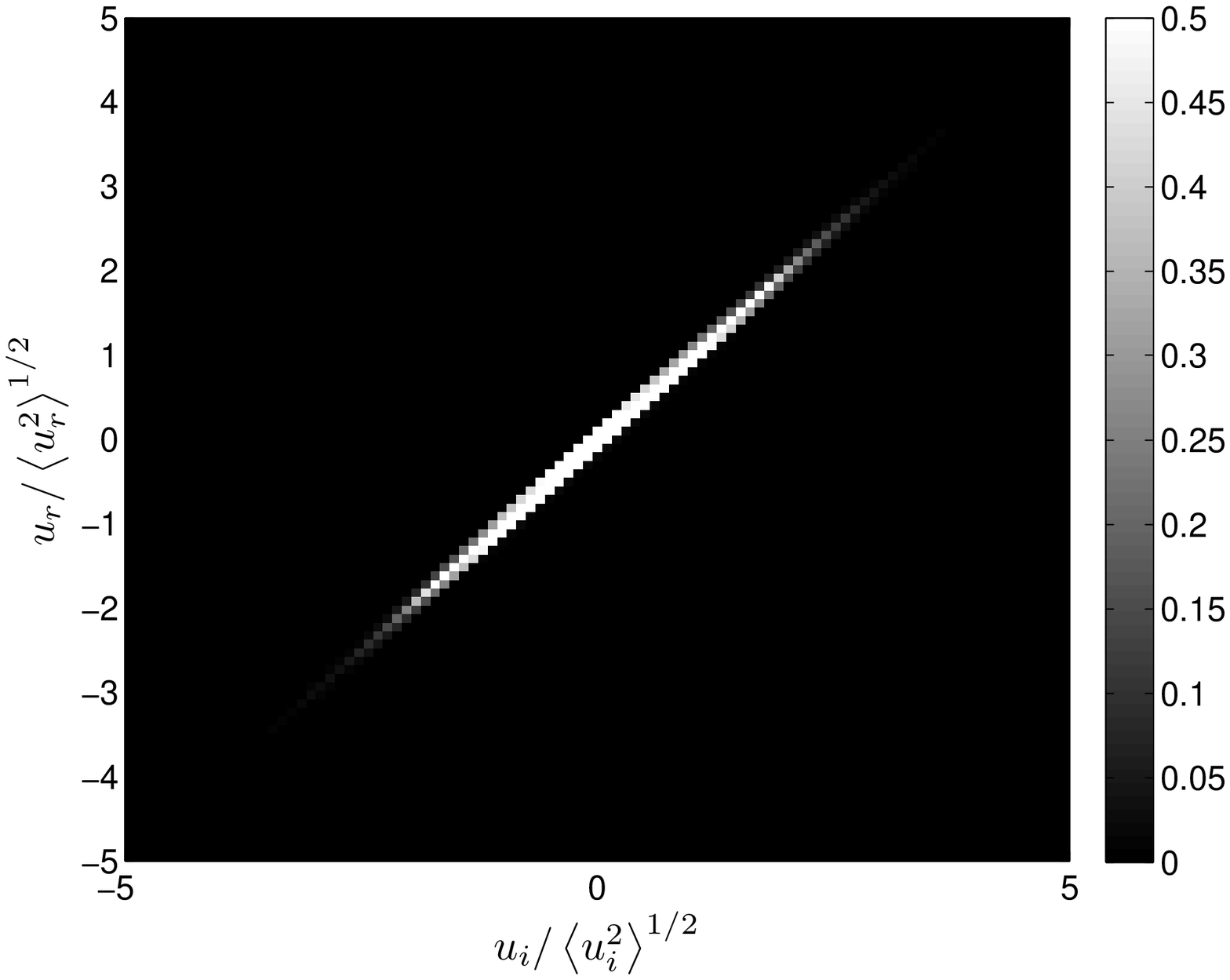}
\label{fig:Jpdff002}}
\end{minipage}
\begin{minipage}[c]{.55\linewidth}
\subfigure[]
{\includegraphics[width=0.9\textwidth]{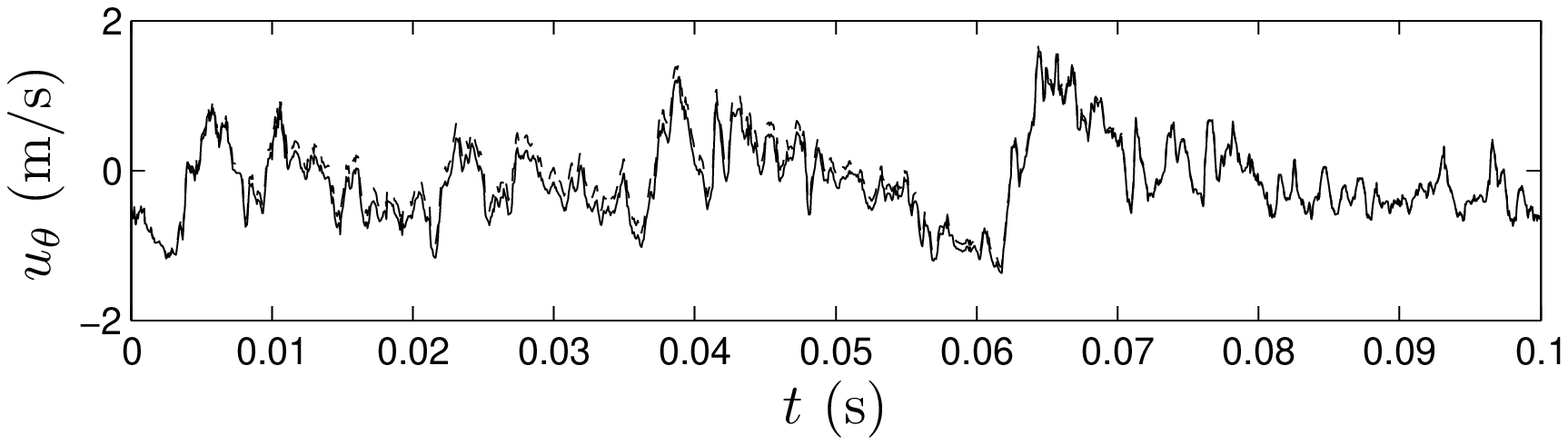}
\label{fig:uf1a2}} 
\subfigure[]
{\includegraphics[width=0.9\textwidth]{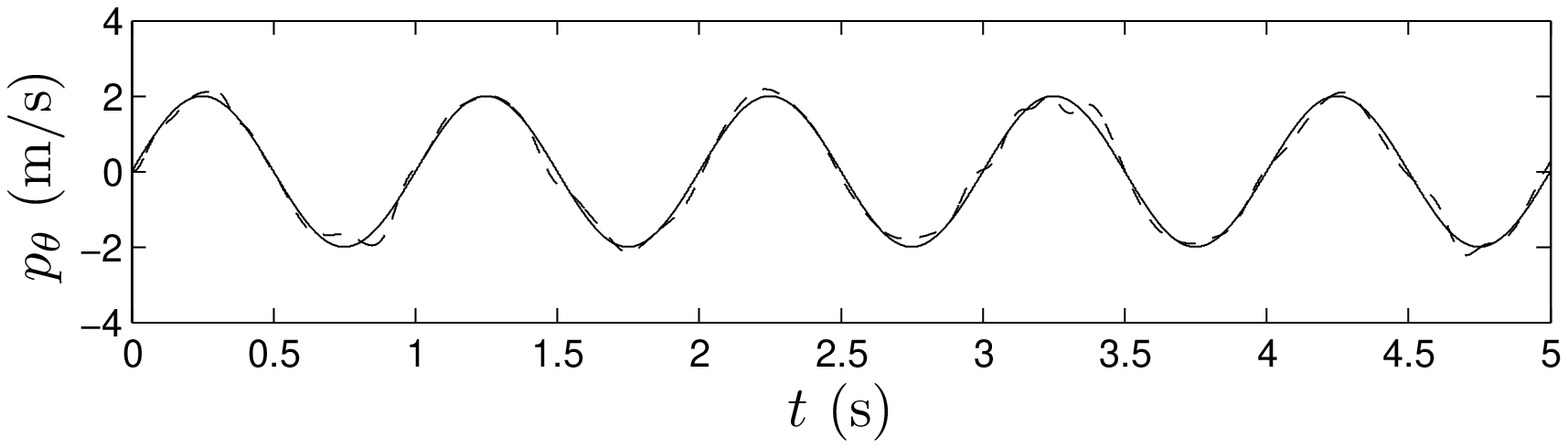}
\label{fig:pf1a2}}
\end{minipage}
\hfill
\begin{minipage}[c]{.41\linewidth}
\subfigure[]
{\includegraphics[width=1\textwidth]{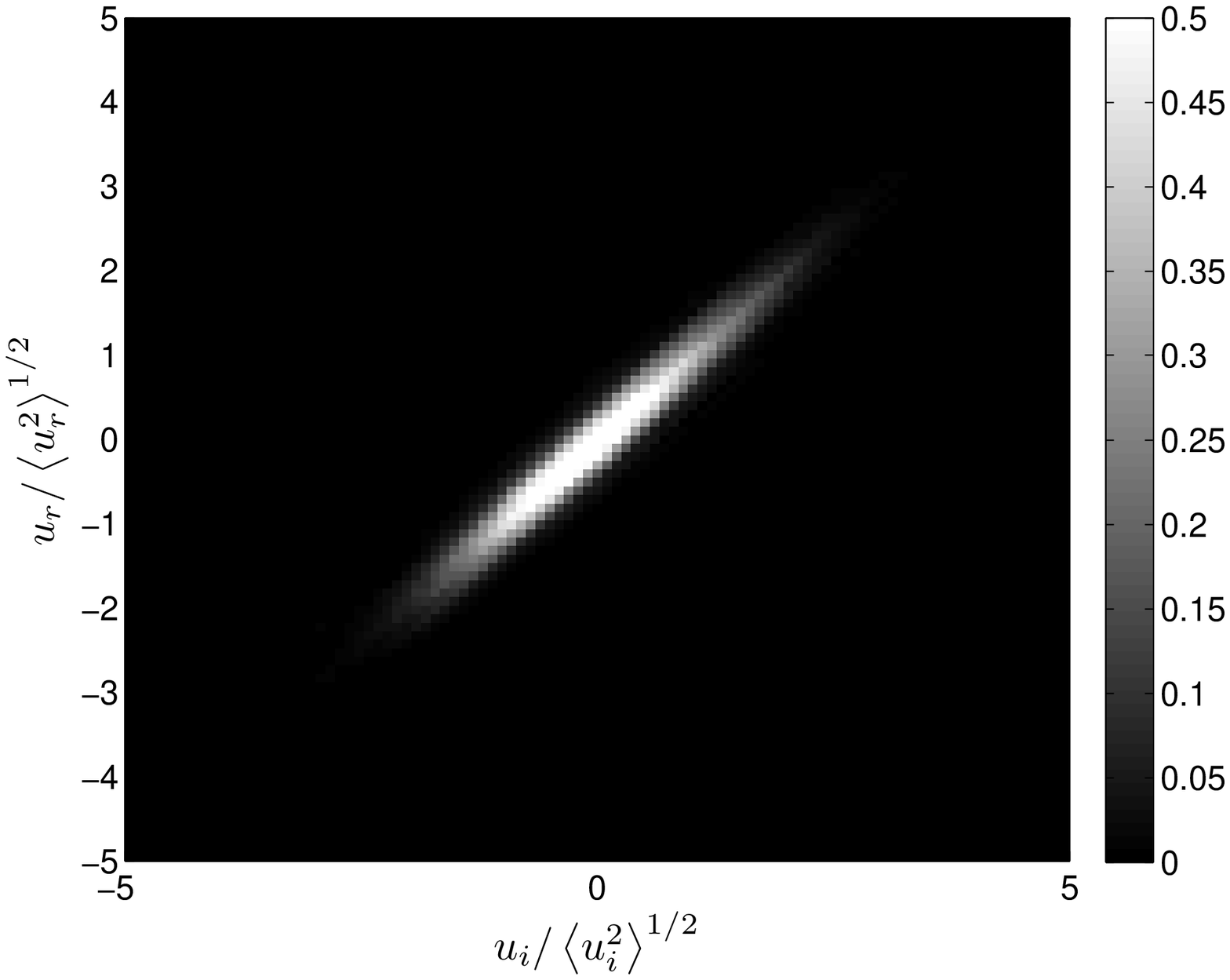}
\label{fig:Jpdff1}}
\end{minipage}
\begin{minipage}[c]{.55\linewidth}
\subfigure[]
{\includegraphics[width=0.9\textwidth]{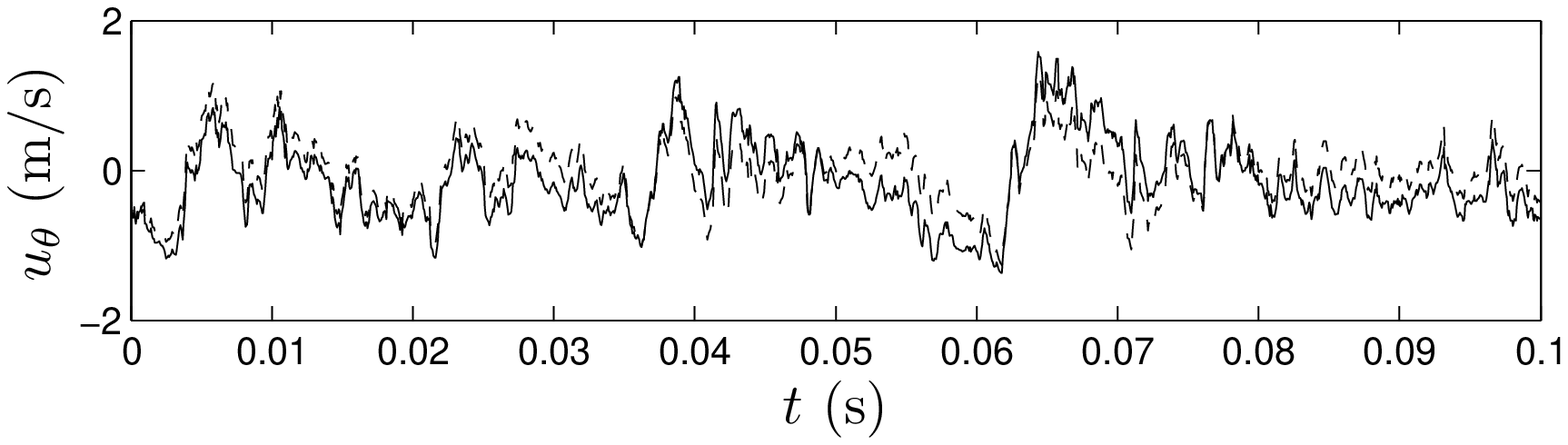}
\label{fig:uf10a2}} 
\subfigure[]
{\includegraphics[width=0.9\textwidth]{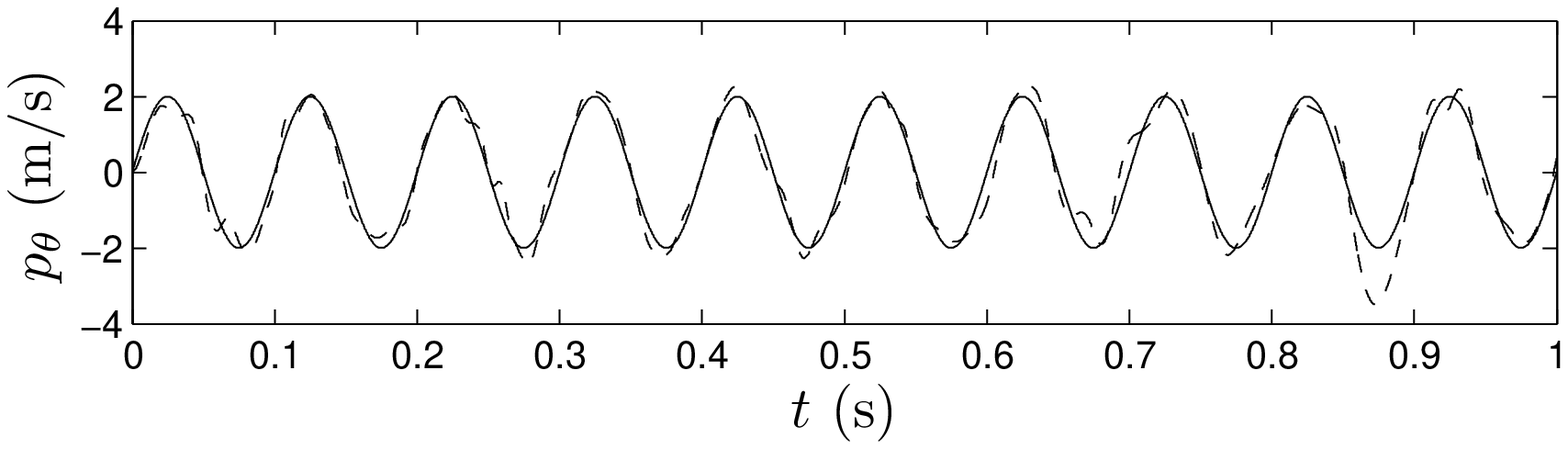}
\label{fig:pf10a2}}
\end{minipage}
\hfill
\begin{minipage}[c]{.41\linewidth}
\subfigure[]
{\includegraphics[width=1\textwidth]{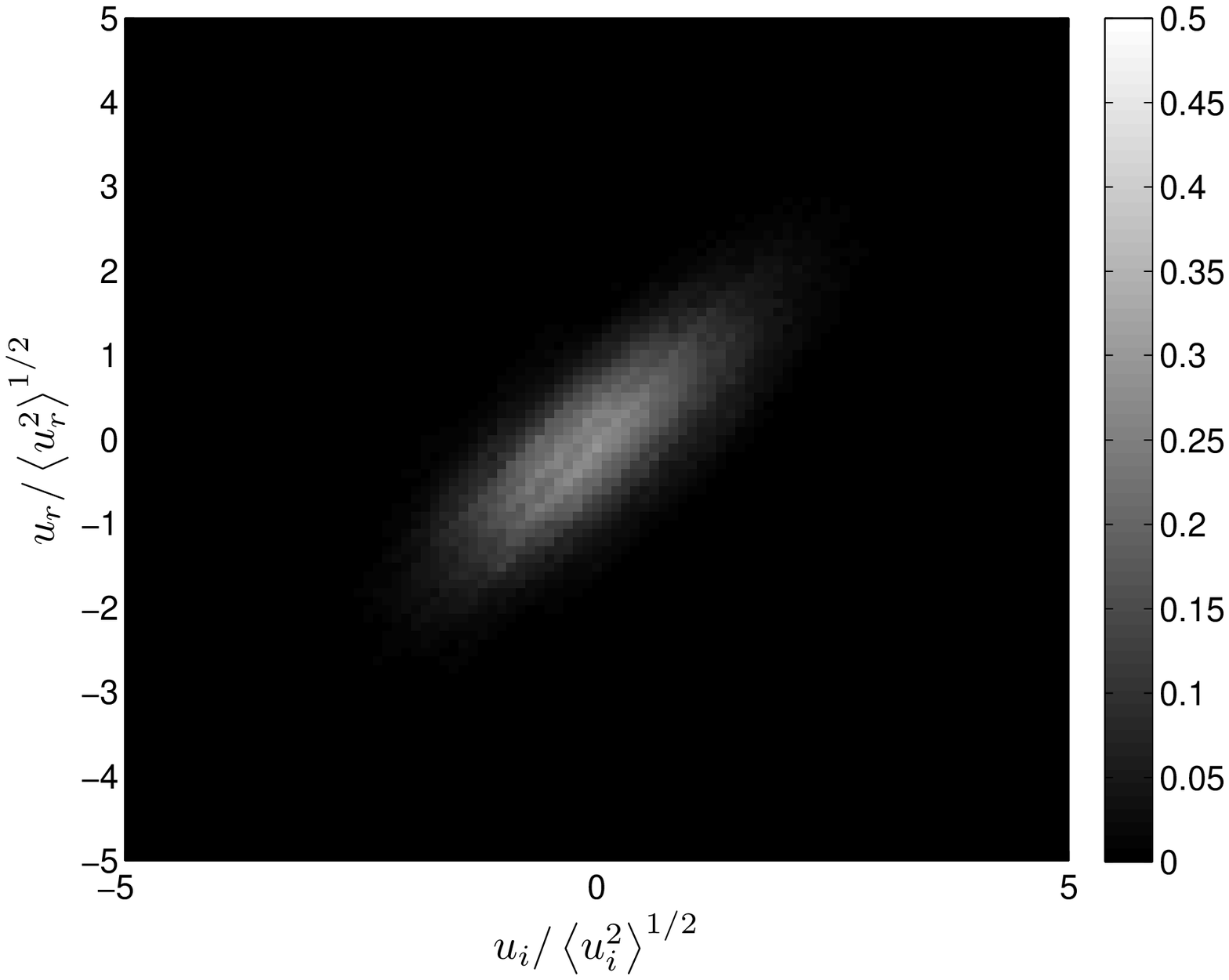}
\label{fig:Jpdff10}}
\end{minipage}
\caption{Recovered signals $u_r(t)$ (dashed lines) vs reference signal
$u_i(t)$ (solid lines)
for three perturbation frequencies ($a_p = 2$ m/s): (a) $f_p = 0.02$Hz,
(d) $f_p = 1$Hz and (g) $f_p = 10$Hz. Comparison between the original
perturbations $p(t)$ and the \textit{a posteriori} estimated
perturbation $p_r(t)$ for three perturbation frequencies
($a_p = 2$ m/s): (b) $f_p = 0.02$Hz, (e) $f_p = 1$Hz and (h) $f_p = 10$Hz.
Joint-PDF between the recovered signal $u_r(t)$
and the reference signal $u_i(t)$ for three specific
"mono-component" perturbations: (a) $f_p = 0.02$Hz,
(b) $f_p = 1$Hz and (c) $f_p = 10$Hz. For each case,
the perturbation amplitude is kept constant $a_p = 2$m/s.}
\end{figure*}

The comparison between the recovered velocity signals $u_r(t)$
and the reference velocity signal $u_i(t)$ is given
in Figs. \ref{fig:uf002a2}, \ref{fig:uf1a2} and
\ref{fig:uf10a2} for $f_p = 0.02$Hz, 1Hz and
10Hz, respectively. For the smallest perturbation frequency,
i.e. $f_p = 0.02$Hz, an impressive agreement between $u_r(t)$ and
$u_i(t)$ is found. This is emphasized by the joint-PDF
between both signals plotted in Fig. \ref{fig:Jpdff002}
which testifies that $u_r(t)$ and $u_i(t)$ are almost indistinguishable.
Furthermore, we draw the reader's attention to the excellent collapse of the
\textit{a posteriori} estimated perturbation $p_r(t)$ on the original
perturbation $p_i(t)$, as shown in Fig. \ref{fig:pf002a2}, even though
the signal duration is much smaller than one period of the perturbation.
It is worth noticing that standard approaches, such as Fourier transform
for instance, would have failed in such accurate deduction because at least  
one period of the perturbation would have been needed. This is an important
feature of the EMD algorithm in order to study unsteady and non-linear
dynamics, in the sense it is a data-driven feature, i.e. without prescribing,
\textit{a priori}, the base functions.

For higher perturbation frequencies,  $u_r(t)$
compares fairly well with $u_i(t)$,
although slight departures are visible (see Figs. \ref{fig:uf1a2} and
\ref{fig:uf10a2}). These discrepancies increase with increasing 
$f_p$, as evidenced by the joint-PDF displayed in
Figs. \ref{fig:Jpdff1} and \ref{fig:Jpdff10} for $f_p$=1Hz
and 10Hz, respectively. Nevertheless, a careful examination of the worst
case, i.e. $f_p = 10$Hz, reveals that the fast dynamics
(high frequency) of the velocity signal is successfully inferred from
the perturbed signal. Therefore, the main observable differences
between $u_r(t)$ and $u_i(t)$ pertain
to their low frequency part. This is emphasized by the comparison
between the \textit{a posteriori} estimated perturbation $p_r(t)$ and the
original perturbation $p_i(t)$ shown in Figs. \ref{fig:pf1a2} and
\ref{fig:pf10a2}.

\subsection{The scale-by-scale turbulent energy}

\begin{figure}[htbp]
\centering
\subfigure[]
{\includegraphics[width=0.65\textwidth]{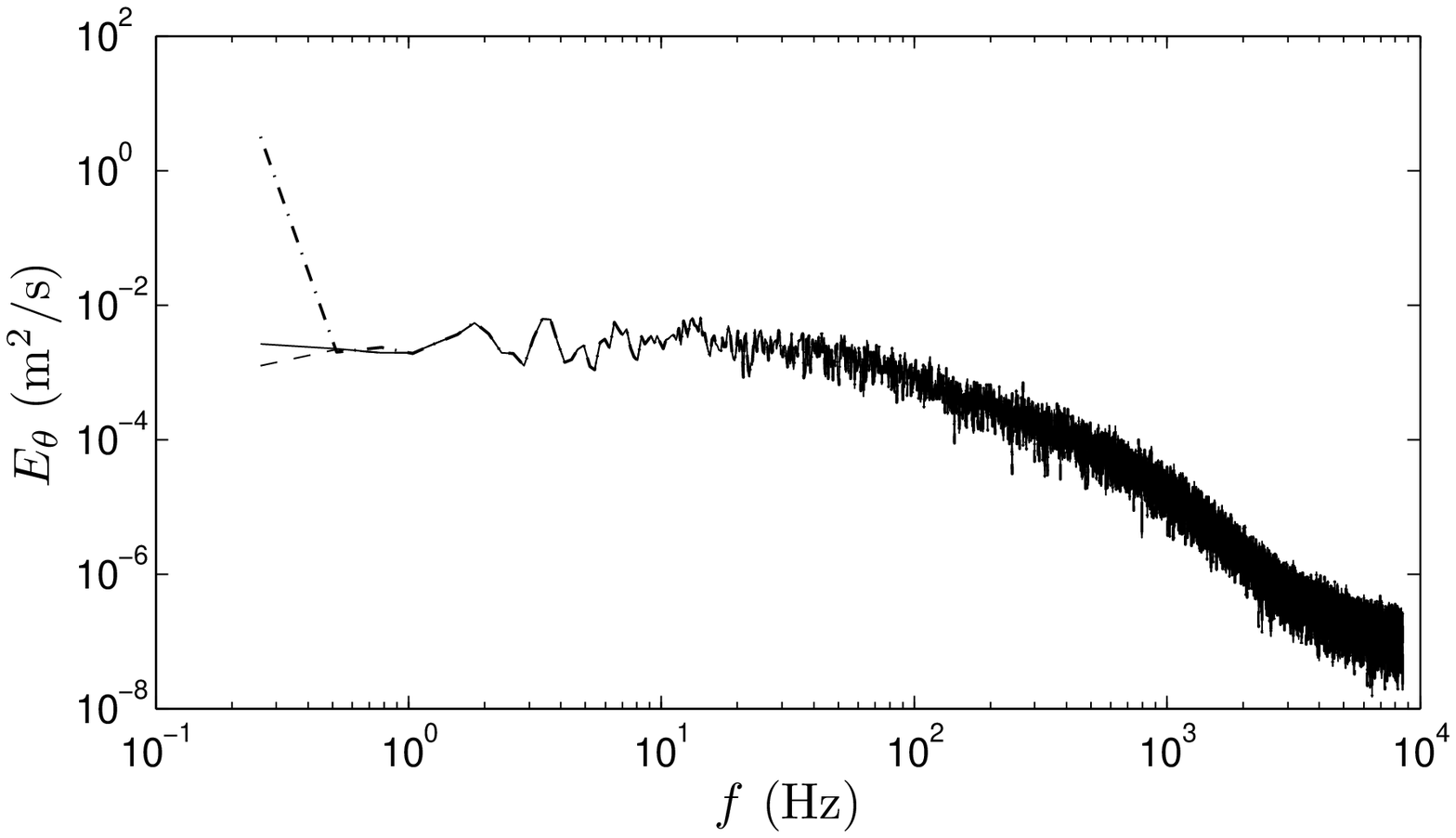}
\label{fig:Specf002a2}}
\vspace{0.1cm}
\subfigure[]
{\includegraphics[width=0.65\textwidth]{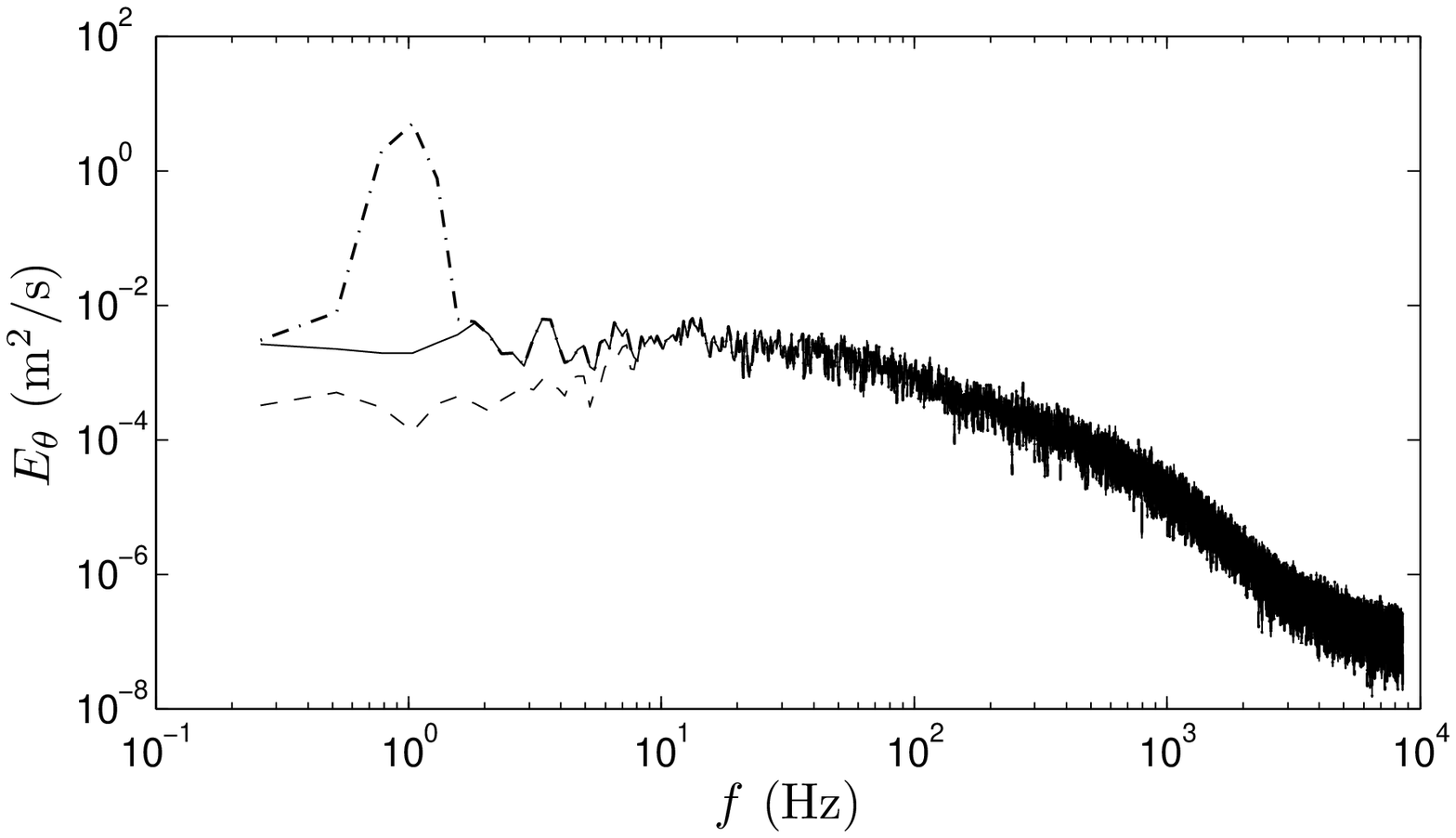}
\label{fig:Specf1a2}}
\vspace{0.1cm}
\subfigure[]
{\includegraphics[width=0.65\textwidth]{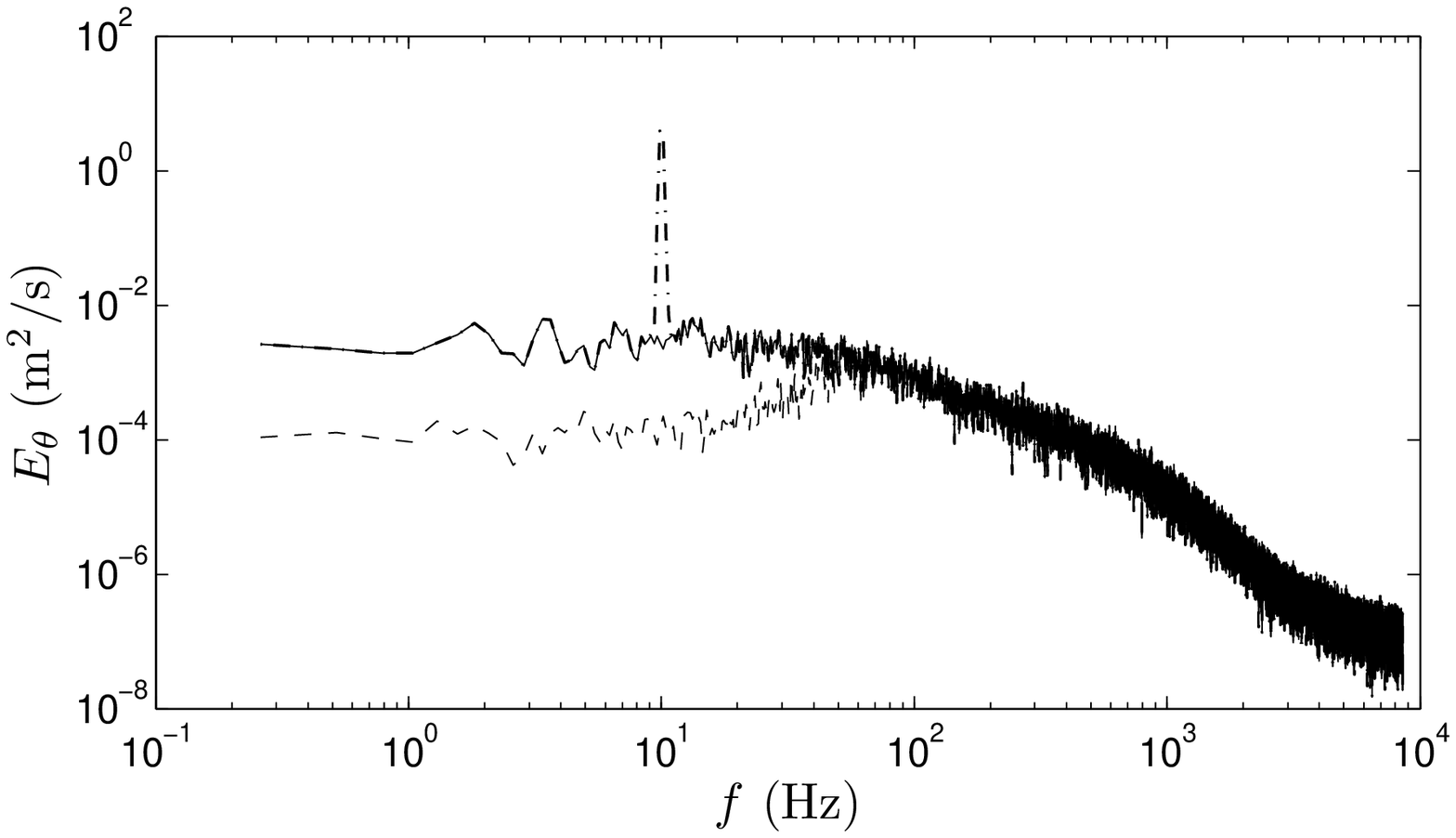}
\label{fig:Specf10a2}}
\hspace{0.1cm}
\caption{1D energy spectra of the reference signal $u_i(t)$ (solid line), perturbed
signals $u_p(t)$ (dotted lines) and recovered signals $u_r(t)$
(dashed lines) for three perturbation frequencies ($a_p = 2$ m/s):
(c) $f_p = 0.02$Hz, (f) $f_p = 1$Hz and (i) $f_p = 10$Hz. }
\end{figure}

A deeper analysis of the recovering procedure can be performed
by comparing the 1D energy spectra $E_\theta$ (defined as
$\left<u_\theta^2\right> = \int_0^\infty E_\theta(f) df$, with $f$ the spectral
frequency), where the subscript $\theta$ refers to either the reference, the
perturbed or the recovered velocity signal. Such a comparison is given in
Figs. \ref{fig:Specf002a2}, \ref{fig:Specf1a2} and \ref{fig:Specf10a2}
for the perturbation frequency $f_p = 0.02$Hz, 1Hz and 10Hz, respectively.
In each case, the perturbation appears as a sharp peak centered on the
frequency $f_p$, even though, as mentioned earlier, for the lowest
perturbation frequency one cannot inferred \textit{a posteriori} the value
of $f_p$ from the spectral analysis. 

One can easily see that both the intermediate inertial range and
the high frequencies of the energy spectra computed from $u_r(t)$
are mostly unaffected by the recovering
procedure, independently of the perturbation frequency. This result
proves that the fast dynamics of the reference signals is extracted with
success from the perturbed velocity signals. The major discrepancies
between $E_r(f)$ and $E_i(f)$ occur in the range $f \leq f_p$. One can
see that the characteristic plateau at low frequency
exhibited by $E_i(f)$ \cite{Pope2000} is underestimated by the
recovering procedure. Both the energy
loss and the range of affected frequencies $f$ increase with
increasing $f_p$. This feature relies on the way our filtering
procedure is designed to reject the high-order polluted IMFs
according to the "resemblance" criterion, as described in Sec.
\ref{sec:EMD}. Indeed, the energy of the rejected IMFs is mostly
concentrated at low frequency (see e.g. Figs.
\ref{fig:SpecIMF5f10a2}-\ref{fig:SpecIMF9f10a2}).

It is not clear whether the intrinsic features of the EMD algorithm,
alone, account for this drawback. Schlotthauer et al. \cite{Schlotthaueretal2009} documented
an extensive study of the influence of parameters such as signal length
and sifting number. They found out that the statistical properties of
the IMFs of a gaussian white noise strongly depend on these parameters.
Another way which can explain this phenomenon stands in the concept
of the local mean value which is defined as the local average of the signal
envelopes. A careful examination reveals that these envelopes can
locally either overshoot or truncate the analyzed signal, altering,
therefore, the quality of the EMD algorithm. Besides these numerical
aspects, one may expect that physical reasons can also account for
the deterioration of the recovering quality with increasing $f_p$.
Indeed, when the perturbation frequency overlaps
the energy-containing frequency range (i.e. the large-scales), one may
expect their respective dynamics to mix, leading to a less efficient
separation by means of the EMD algorithm. Rilling and Flandrin \cite{RillingFlandrin2008} 
reported that both modes of a "two-components"
signal can be sometimes unfruitfully separated by the EMD method, depending
on the frequency ratio of both modes as well as their relative amplitude.

Nevertheless, it is worth noting that even though a significant part of
the energy at low frequency is lost during the rejection procedure, this
is much less severe than standard high-pass filtering. Furthermore, one can argue
that band-pass filtering could have performed more efficiently but
such approach implies the unwanted frequency band to be clearly
identifiable. This point is of particular importance, especially
for the case of "multi-component" perturbation which is investigated
at the end of this section.

As pointed out by Huang et al. \cite{Huangetal1998}, the use of the Fourier
transform (and related tools) to study non-stationary physics is
meaningless. The scale-by-scale investigation of the turbulence by means of the
structure functions is an alternative and complementary way to the
spectral analysis. Structure functions are defined as the statistical
moments of the velocity increments, i.e. the velocity difference
between two points separated by a time increment
$\tau$ \cite{Frisch1995}, which, in this study, are restricted
to the streamwise velocity increment 

\begin{equation}
\delta u_\theta(\tau) = u_\theta(t+\tau) - u_\theta(t).
\label{eq:increment}
\end{equation}

\begin{figure}[htbp]
\centering
\subfigure[]
{\includegraphics[width=0.65\textwidth]{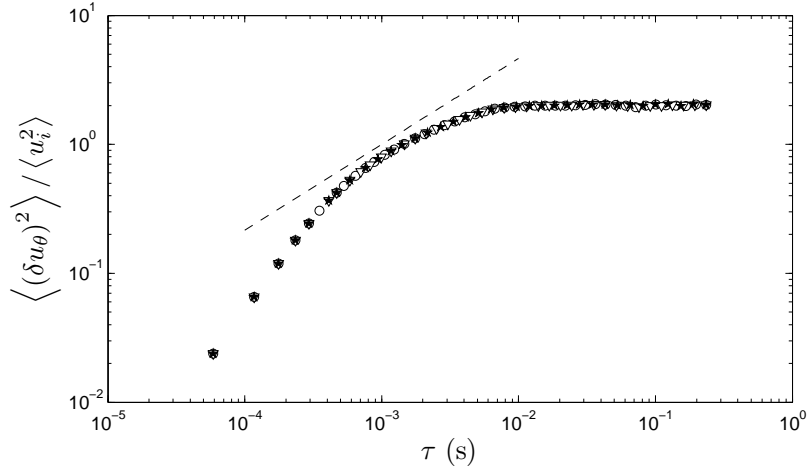}
\label{fig:d2uf002}}
\vspace{0.1cm}
\subfigure[]
{\includegraphics[width=0.65\textwidth]{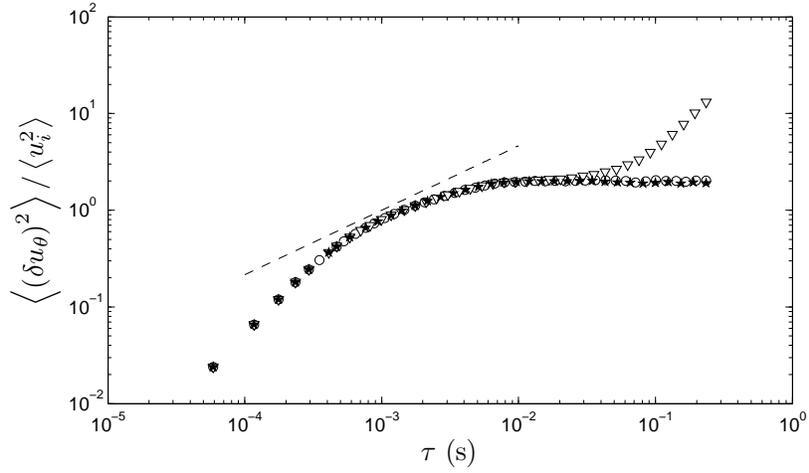}
\label{fig:d2uf1}}
\vspace{0.1cm}
\subfigure[]
{\includegraphics[width=0.65\textwidth]{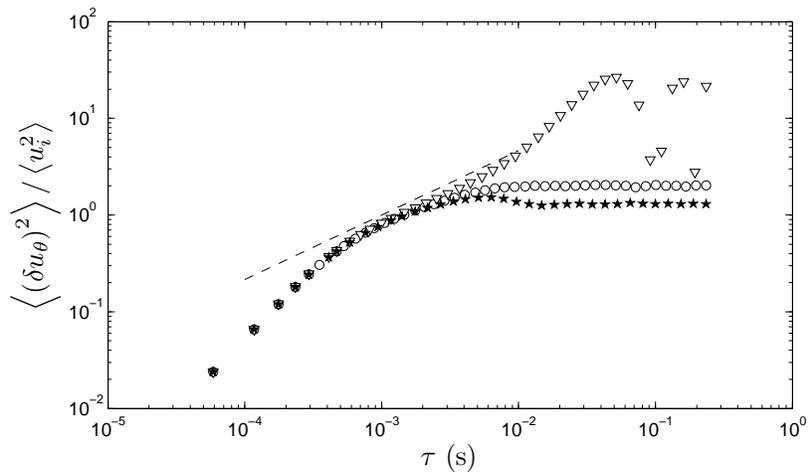}
\label{fig:d2uf10}}
\caption{Second-order structure functions of the reference
($\circ$), perturbed ($\triangledown$) and recovered ($\star$)
velocity signals for (a) $f_p = 0.02$Hz, (b) $f_p = 1$Hz
and (c) $f_p = 10$Hz. For each perturbed signal, the perturbation
amplitude is constant ($a_p = 2$m/s). The scaling law $\tau^{2/3}$
is represented by the dash lines.}
\end{figure}

The dimensionless second-order structure function
$\left<(\delta u_\theta)^2\right> / \left<u_i^2\right>$
computed from the reference, perturbed and recovered velocity signals are plotted
in Figs. \ref{fig:d2uf002}, \ref{fig:d2uf1} and \ref{fig:d2uf10} for
 $f_p = $0.02Hz, 1Hz and 10Hz, respectively.

Physically, $\left<(\delta u_\theta)^2\right>$ represents the energy of
a scale $\ell$, where the spatial separation is deduced from the time
lag $\tau$ by means of the Taylor's hypothesis, i.e. $\ell = -U \tau$.
One can show that $<\left(\delta u_\theta\right)^2>$ tends towards
$2 \left<u_\theta^2\right>$ for large time (or space according to the
Taylor's hypothesis) separation, i.e. when $\tau \rightarrow \infty$.
This expectation is well supported by the results computed from the
reference velocity signal. Moreover,
a narrow inertial range can be identified by the scaling law 
$\left<(\delta u_\theta)^2\right> \propto \tau^{2/3}$, equivalent to
the well-known $-5/3$ scaling law in spectral space
\cite{TennekesLumley1972, Pope2000}.

Obviously, the
long-period flapping, simulated by the numerical perturbation, has
a significant effect at large $\tau$, especially for the two highest
perturbation frequencies (see Figs. \ref{fig:d2uf1} and
\ref{fig:d2uf10}). Furthermore, in the case of $f_p = 10$Hz, the
perturbation affects also the beginning of the inertial range.

The results obtained for the recovered velocity signals clearly evidence
the high quality of the recovering procedure, in particular
for $f_p = 0.02$Hz and 1Hz. In both cases, an impressive collapse onto
the reference second-order structure function is found. These results indicate that,
for $f_p = 0.02$Hz and 1Hz, the amount of turbulent energy contained in
the rejected IMFs is almost negligible in comparison with the turbulent
kinetic energy $\left<u_i^2\right>$. For the highest
frequency perturbation, i.e. $f_p=$10Hz, although the plateau value is
underestimated after the recovering procedure (notice that less than half
of the IMFs are kept according to Fig. \ref{fig:Rf10}),
the influence of the perturbation is successfully annihilated, particularly
regarding the inertial range.

\subsection{The turbulence properties}

The quantitative assessment of the recovering quality can be done by
comparing the statistical properties of both the reference and the
recovered velocity signals, at large- and small-scales. The large-scales
are featured by both their energy, i.e. $\left<u_\theta^2\right>$, and the integral
length-scale

\begin{equation}
L_\theta = \frac{U}{u_\theta'^2} \int_0^\infty \left\langle u_\theta(t)
u_\theta(t + \tau)\right\rangle d\tau.
\label{eq:L}
\end{equation}

The small-scales can be described through the Taylor micro-scale

\begin{equation}
\lambda_\theta = U \sqrt{\frac{\left<u_\theta^2\right>}{\left< \left(\frac{\partial
u_\theta}{\partial t}\right)^2\right>}},
\label{eq:lambda}
\end{equation}

and the Kolmogorov scale

\begin{equation}
\eta_\theta = \left(\frac{\nu^3}{\epsilon_\theta}\right)^{1/4},
\label{eq:eta}
\end{equation}

where $\epsilon_\theta$ ($\equiv \frac{15 \nu}{U^2} \left\langle \left(\frac{\partial
u_\theta}{\partial t}\right)^2\right\rangle$ under isotropic assumption) is the
turbulent kinetic energy dissipation rate.

\begin{figure}[htbp]
\centering
\includegraphics[width=0.48\textwidth]{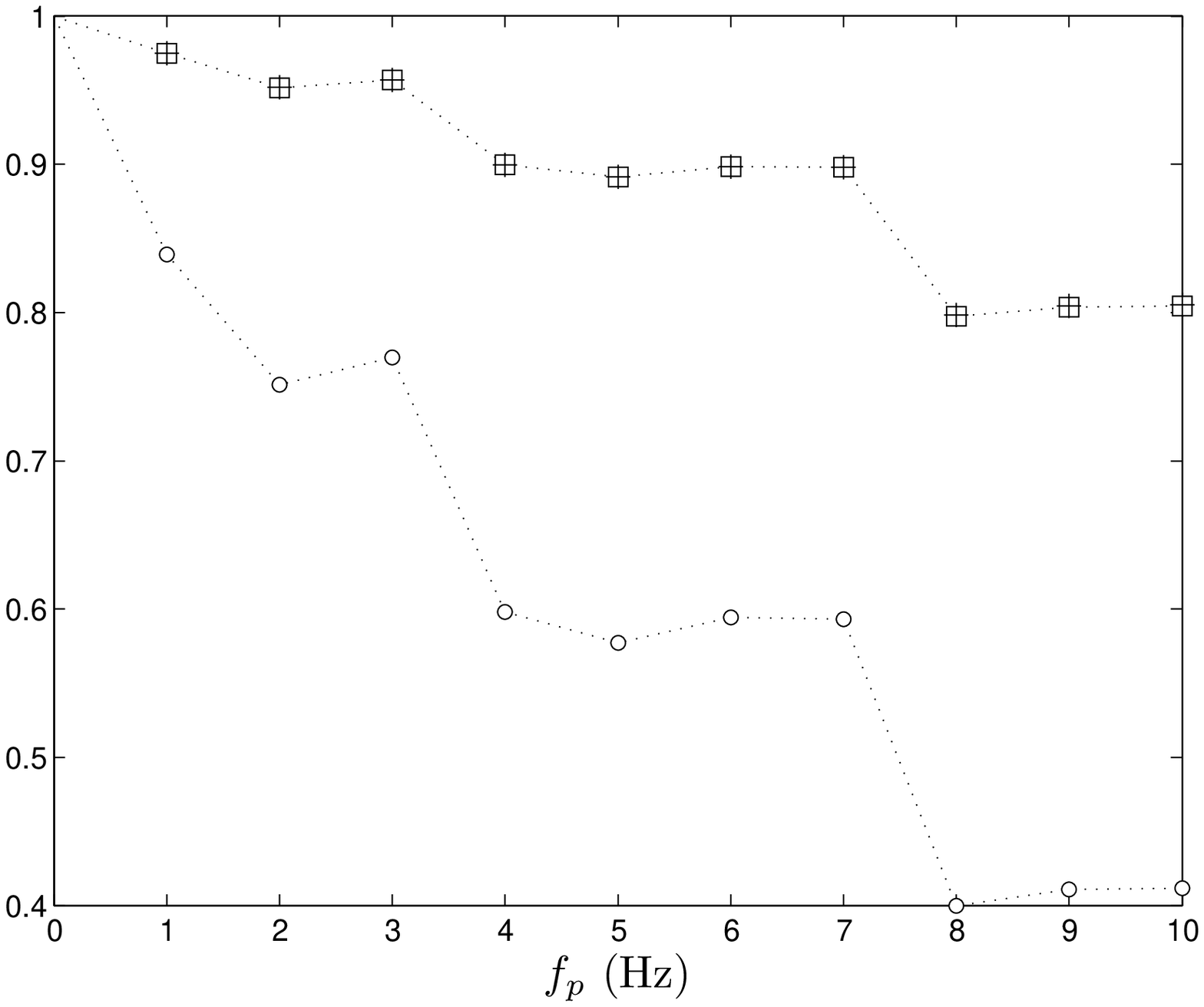}
\caption{Variation of $(\left<u_r^2\right>/\left<u_i^2\right>)^{1/2}$
($\square$), $\lambda_r / \lambda_i$ ($+$)
and $L_r/L_i$ ($\circ$) against $f_p$.}
\label{fig:Ratio}
\end{figure}

The ratios $(\left<u_r^2\right>/\left<u_i^2\right>)^{1/2}$, $\lambda_r / \lambda_i$
and $L_r/L_i$ are plotted in Fig. \ref{fig:Ratio} with respect to $f_p$. One can
see that the differences between the recovered and the reference signals
increase with increasing $f_p$. The staircase-like shape of the curves is
related to the rejection method which acts as a discrete filter. Each step
corresponds to the rejection of an additional IMF. 

Over the whole range of $f_p$ tested here, the recovering of
$\left<u_r^2\right>^{1/2}$ is better than 80\%. Furthermore, we
have shown, hereinbefore, that the high frequency dynamics is
unaffected by the recovering algorithm implying, therefore, that the
statistics of the derivatives are almost unchanged, i.e.
$\epsilon_r \approx \epsilon_i$ and $\eta_r \approx \eta_i$.
Consequently, according to Eq. (\ref{eq:lambda}), the uncertainties in the
evaluation of $\lambda_r$ relies only on those of $\left<u_r^2\right>^{1/2}$,
explaining why $(\left<u_r^2\right>/\left<u_i^2\right>)^{1/2}$ and
$\lambda_r/\lambda_i$ are almost indistinguishable.

The largest discrepancies are observed for the integral length-scale which
is significantly underestimated, by about 60\% for the highest $f_p$. This
result can be explained by reformulating Eq. (\ref{eq:L}) under the following
form \cite{Pope2000} 

\begin{equation}
L_\theta = \frac{U E_\theta(f=0)}{4 \left<u_\theta^2\right>}.
\label{eq:L2}
\end{equation}

As shown previously (see e.g. Fig. \ref{fig:Specf10a2}), the term $E_r(f=0)$
is dramatically affected by the rejection procedure, even though it would
have been worst with a classical high-pass filter. The strong decrease in  
$E_r(f=0)$ accounts for the underestimation of $L_r$. This is an important
difference with $\left<u_r^2\right>$ which is computed from the integration of $E_r$
and is, therefore, less affected by the energy loss at low frequency.

An important issue, linked to the interaction between large- and small scales,
arises from these results. Investigating the features of the zero-crossing of
the velocity fluctuation in various experiments, \cite{MazellierVassilicos2008}
found out an universal relationship between the dissipation constant and the
large-scale flow topology. Consequently, one may assume that the alteration of
the large-scale properties (highlighted by the underestimation of the integral
length-scale) might also modify directly or not the inertial range and/or the
small-scale properties. This issue might actually be addressed through the
investigation of the third-order structure function $\left<(\delta u_\theta)^3\right>$
which represents the energy transfer at a given scale. In the following,
for sake of simplicity, we consider the spatial separation $\ell$ instead of
the time separation $\tau$. For high enough Reynolds number, the dimensionless
third-order structure function $S_{3 \theta} = - \left<(\delta u_\theta)^3\right> /
(\epsilon_\theta \ell)$ is expected to reach a constant value in the inertial
range \cite{Frisch1995}

\begin{equation}
S_{3 \theta} = \frac{4}{5}.
\label{eq:S3}
\end{equation}

Using independent approaches, Qian \cite{Qian1999} and afterwards Lundgren \cite{Lundgren2002}
proposed corrections of Eq. (\ref{eq:S3}) accounting for the effects of
the finite Reynolds number in the framework of freely decaying turbulence.
These corrections can be expressed under the following form

\begin{equation}
S_{3\theta} = \frac{4}{5} - C_1 \left(\frac{\ell}{\ell_o}\right)^\alpha
- C_2 \left(\frac{\ell}{\ell_i}\right)^\beta,
\label{eq:S32}
\end{equation}

where $\ell_o$ and $\ell_i$ represent an outer scale (i.e. large scale)
and an inner scale (i.e. viscous scale), respectively. For decaying turbulence,
the exponents $\alpha$, $\beta$ and the numerical constants $C_1$, $C_2$ are
equal to $2/3$, $-4/3$ and $3.34/\sqrt{15}$, $8$, respectively (see
Eq. (4) given by Gagne et al. \cite{Gagneetal2004}). This relationship
accounts for the influence of the large- and small-scales on the energy
transfer at a scale $\ell$. Taking $\ell_o = L_\theta$ and $\ell_i =
\eta_\theta$, the maximum value reached by $S_{3\theta}$ (defined by
$dS_{3\theta}/d\ell = 0$), according to Eq. (\ref{eq:S32}), is

\begin{equation}
S_{3\theta}^{max} = \frac{4}{5} - \gamma \left(\frac{\eta_\theta}
{L_\theta}\right)^{4/9},
\label{eq:S3max}
\end{equation}

with $\gamma = 3/2 C_1 (2 C_2 / C_1)^{1/3} \approx 3.42$. 
Gagne et al. \cite{Gagneetal2004} reported a good agreement between this
prediction and experimental data obtained in various turbulent flows. These
authors proposed a Reynolds-correction to compensate for the deviation
at large- and small-scale. Even though the data collapsing was
significantly improved, the use of these corrections is beyond the scope of the
present study where we essentially focused on the maximum
value of $S_{3\theta}$. Indeed, one of the main interests of Eq.
(\ref{eq:S3max}) is to provide an alternative way to Eq. (\ref{eq:L})
to estimate $L_\theta$. 
 
\begin{figure}[htbp]
\centering
\includegraphics[width=0.48\textwidth]{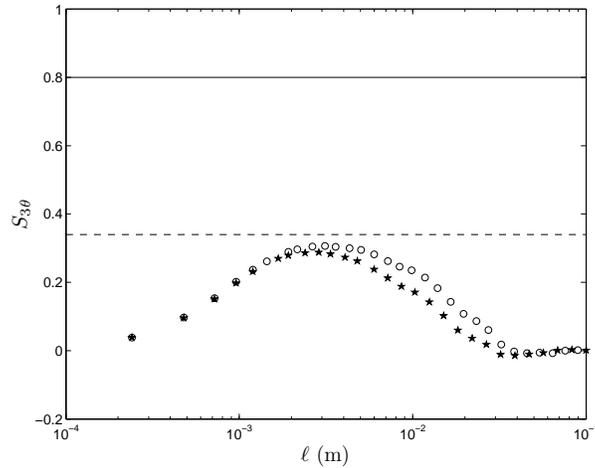}
\caption{Evolution of $S_{3i}$ ($\circ$) and $S_{3r}$ ($\star$) with respect
to $\ell$ (inferred from the time separation $\tau$ by means of the Taylor's
hypothesis). The solid line symbolizes
the "$4/5$" law \cite{Frisch1995}. The dashed line symbolizes the
Lundgren's prediction \cite{Lundgren2002} for the reference signal
according to Eq. (\ref{eq:S3max}). These results have been obtained for a
"mono-component" perturbation with $f_p=10$Hz and $a_p=2$m/s.}
\label{fig:S3}
\end{figure}

The evolution of $S_{3\theta}$ computed from $u_i(t)$ and $u_r(t)$
is plotted in Fig. \ref{fig:S3} against $\ell$. It is worth noting that no significant
difference between $S_{3i}$ and $S_{3r}$ has been observed for 
$f_p = 0.02$Hz and 1Hz. Therefore, only the results obtained for the
worst case, i.e. $f_p = 10$Hz are reported in Fig. \ref{fig:S3}.
The maximum value of $S_{3i}$ according to Eq. (\ref{eq:S3max})
is displayed in this plot. To emphasize the relevance of the Lundgren's
corrections, the "4/5" law is also plotted. One can clearly see that the Lundgren's
prediction is a good approximation of $S_{3i}^{max}$, even though the
theoretical inertial range plateau is slightly overestimated. The discrepancies
between the expected and the experimental values of $S_{3i}^{max}$ may
be accounted for the non-stationarity feature of decaying turbulence
as pointed out by Danaila et al. \cite{Danailaetal1999}.

The comparison between $S_{3i}$ and $S_{3r}$ shows that the influence of
the recovering procedure is noticeable at large-scale and in the inertial
range as well. The decrease of the Reynolds number $Re_\lambda$, due to the
underestimation of both the turbulent kinetic energy and the Taylor
micro-scale, may account for these results (see e.g. Fig. 1 of
\cite{Gagneetal2004}). Thanks to the measurement of $S_{3\theta}^{max}$,
the integral length-scale of the recovered signal, $L_r$, can be
estimated according to Eq. (\ref{eq:S3max}). In that case, the ratio
$L_r / L_i$ is found to be better than 90\%.
This is an impressive improvement compared to the values reported
in Fig. \ref{fig:Ratio}. It seems, therefore, that even though the
rejection method can alter the statistical properties of
the turbulent signal, the underlying physics is still conserved during
the recovering procedure. This may be accounted for the specific features
of the IMFs which span a broad band of frequencies (or wavenumbers)
as evidenced in Figs. \ref{fig:SpecIMF5f10a2}-\ref{fig:SpecIMF9f10a2}.
These plots support the idea that each IMF contains a more or less
important part of large- and small-scale information.

\subsection{The non-stationary flapping}

We close this section by assessing the performances of the recovering
procedure on a time-dependent perturbation mimicking the non-stationary
feature often encountered in engineering applications (e.g. unsteady
injection in engines). For that purpose, the perturbation added to the
reference signal is a linear chirp, i.e. $f_p(t) = f_0 (t/t_m)$ with
$f_0 = 10$Hz ($a_p = 2$m/s). The perturbation frequency sweeps the entire
range previously reported for the "mono-component" perturbation.

\begin{figure}[htbp]
\centering
\subfigure[]
{\includegraphics[width=0.80\textwidth]{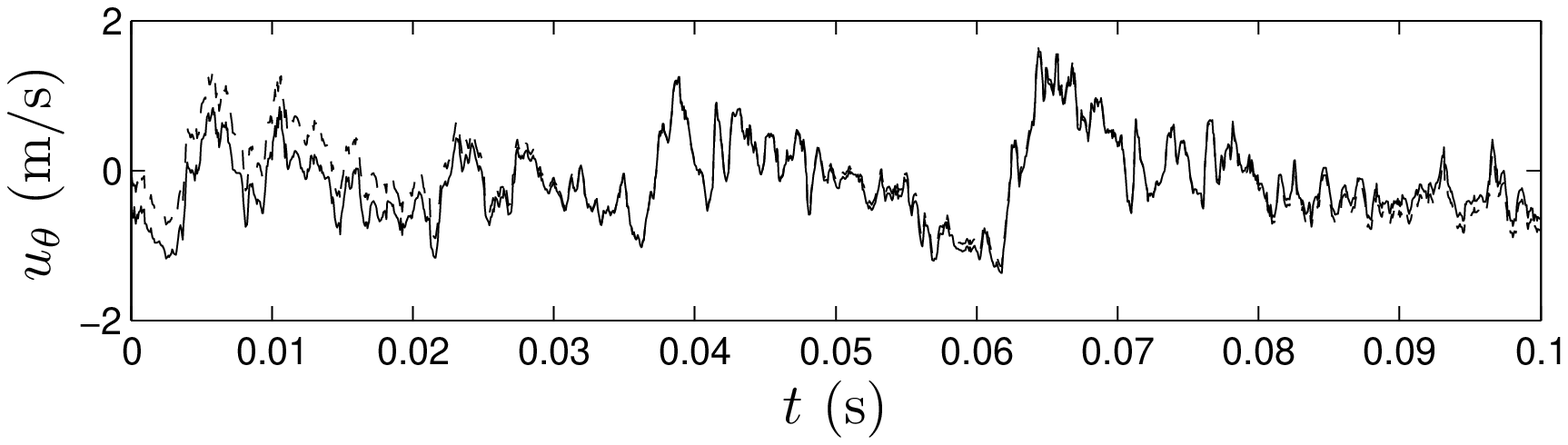}
\label{fig:ufchirplin}}
\vspace{0.1cm}
\subfigure[]
{\includegraphics[width=0.80\textwidth]{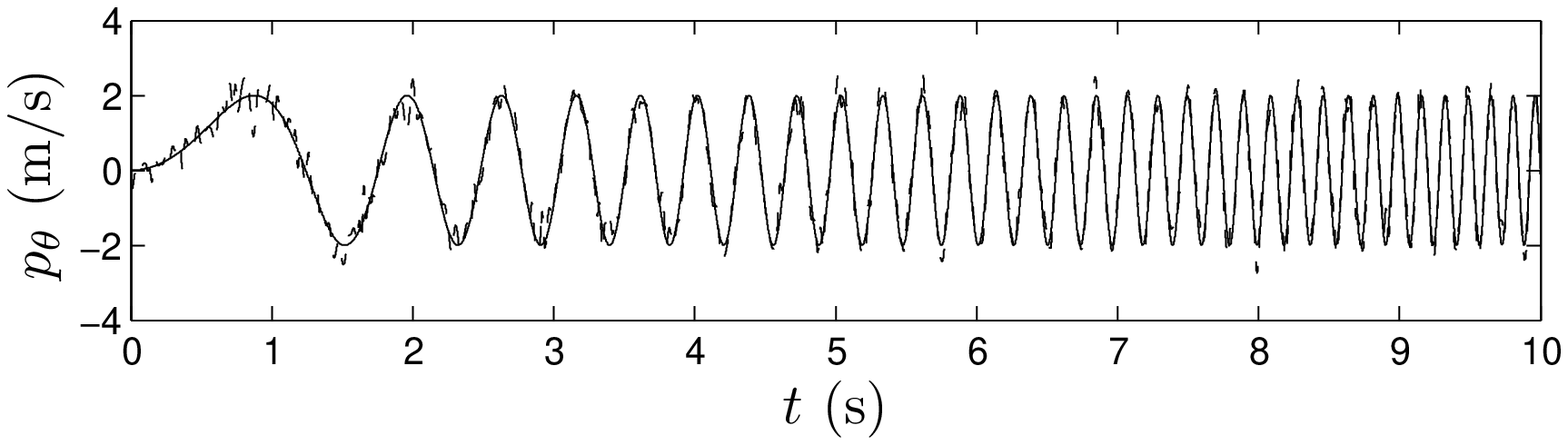}
\label{fig:pfchirplin}}
\caption{(a) Recovered velocity signal $u_r$ (dash line) compared
with the reference velocity signal $u_i$ (solid line). (b) \textit{A
posteriori} estimated perturbation $p_r$ (dash line) compared
with the original perturbation $p_i$ (solid line). These results have
been obtained in the case of the
time-dependent perturbation (linear chirp).}
\end{figure}

The recovered velocity signal $u_r$ and the \textit{a posteriori}
estimated perturbation $p_r$ are compared with their reference
counterparts in Figs. \ref{fig:ufchirplin} and \ref{fig:pfchirplin},
respectively.

The excellent agreement between the recovered and the original signals
testifies to the efficiency of the EMD algorithm in separating both
contributions. This claim applies especially regarding the dynamics
of both the velocity and the perturbation, albeit in the frequency
range investigated in this study. Even though a weak scatter is noticeable
at low perturbation frequency (see Fig. \ref{fig:pfchirplin}), the
rejection procedure captures successfully the frequency shift and the
perturbation amplitude as well.

\begin{figure}[htbp]
\centering
\subfigure[]
{\includegraphics[width=0.45\textwidth]{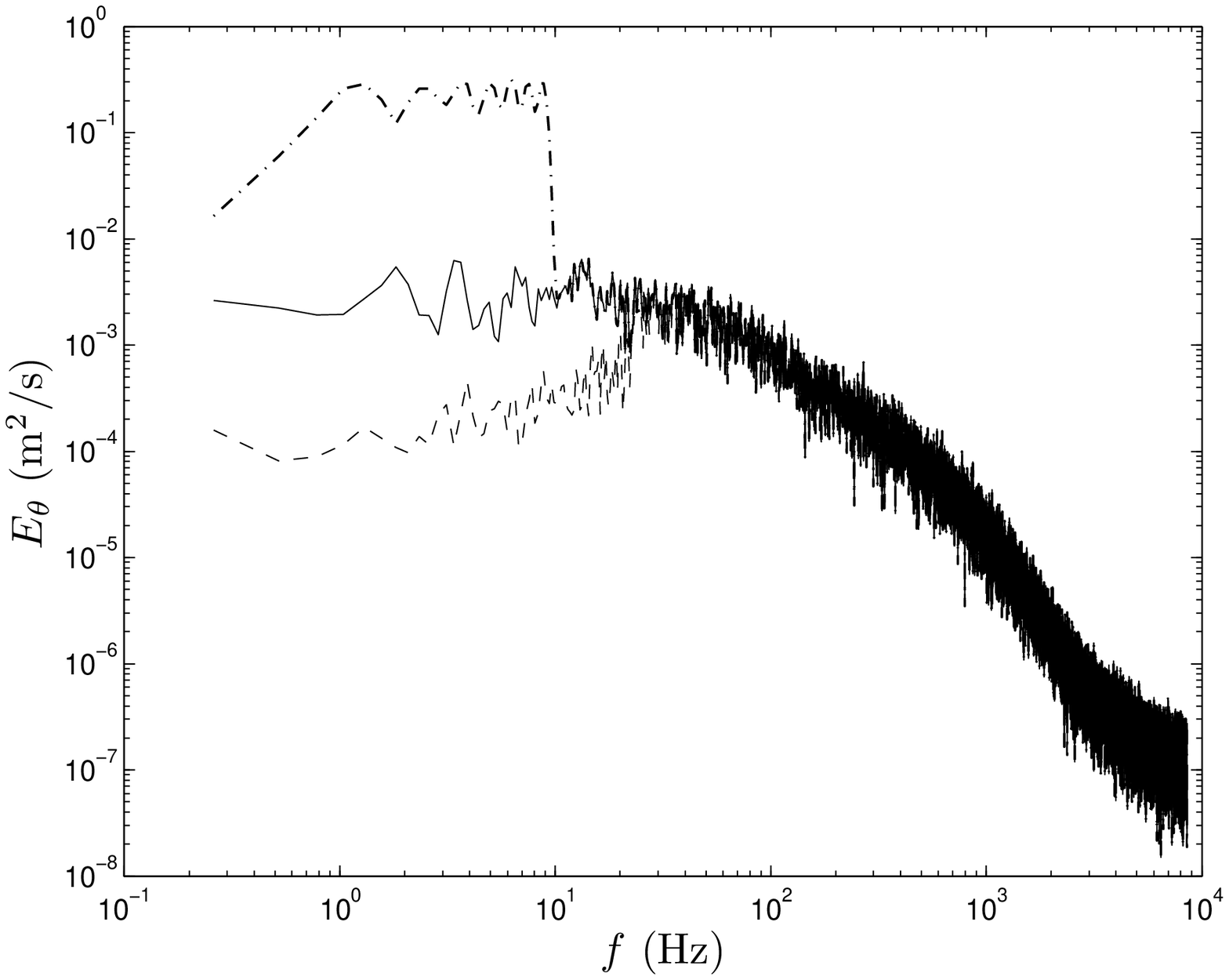}
\label{fig:Specfchirplin}}
\vspace{0.1cm}
\subfigure[]
{\includegraphics[width=0.45\textwidth]{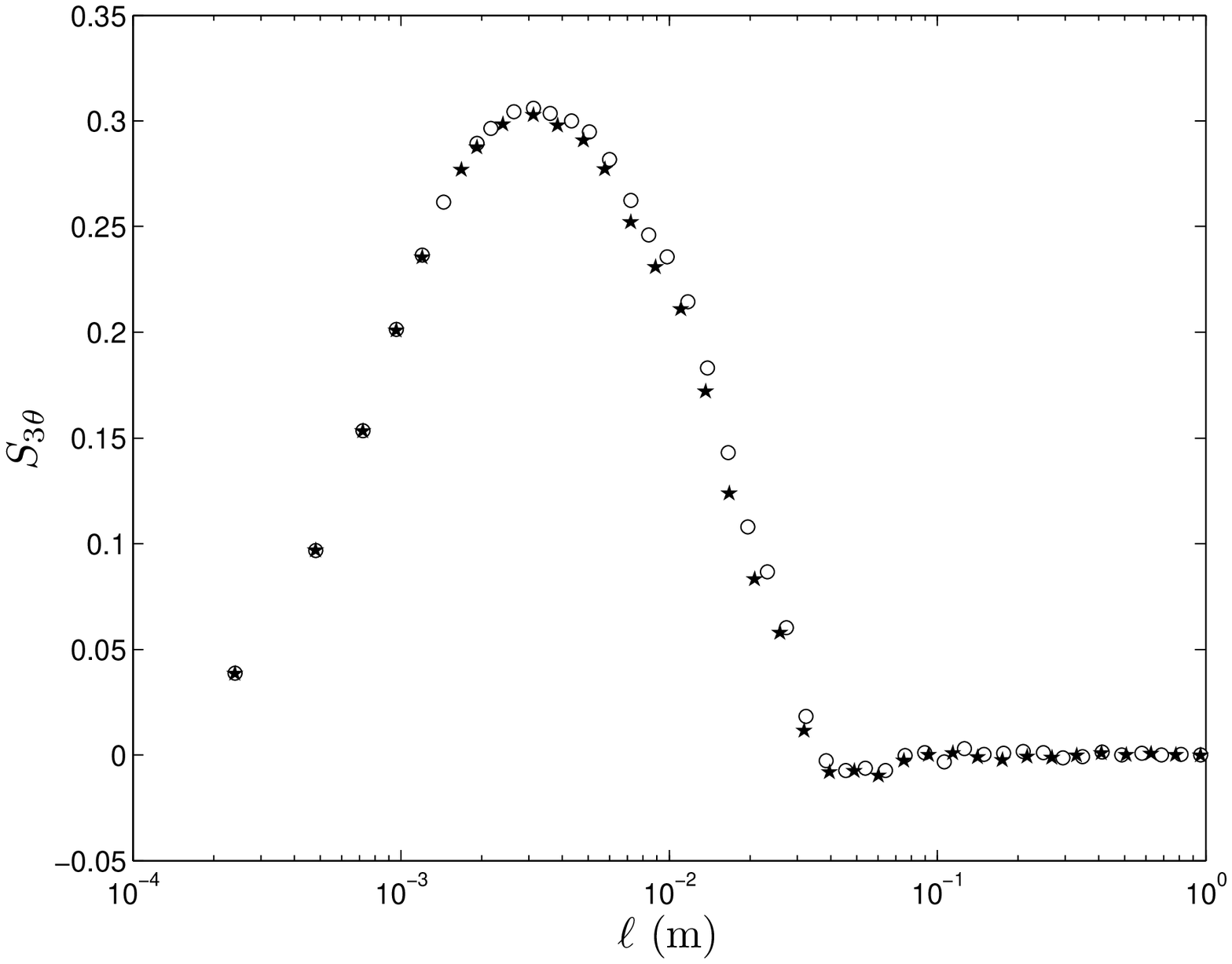}
\label{fig:S3chirp}}
\caption{(a) 1D energy spectra computed from $u_i$ (solid line), $u_p$
(broken line with dots) and $u_r$ (dash line). (b) Dimensionless third-order
structure function computed from $u_i$ ($\circ$) and $u_r$ ($\star$)
against the separation scale $\ell$ (inferred via the Taylor's
hypothesis). These results have been obtained in the case of the
time-dependent perturbation (linear chirp).}
\end{figure}

The effect of the rejection procedure is highlighted by the energy spectra
plotted in Fig. \ref{fig:Specfchirplin}. As expected, the perturbation
(linear chirp) adds a large amount of energy over a broad range of frequency.
One can clearly see that the recovering algorithm annihilates the energy
contained in the numerical perturbation, even though the turbulent energy
$E_r$ is underestimated in the low frequency range as observed in the
case of the "mono-component" perturbation. However, the energy
transfer through scales is much less altered by the rejection process
as shown by the dimensionless third-order structure function $S_{3\theta}$
displayed in Fig. \ref{fig:S3chirp}. This can be partly explained by the fact that
the highest frequencies of the perturbation are less energetic than
in the investigation of the "mono-component" perturbation.

\section{Conclusion}

The performances of a new data analysis method, namely the Empirical Mode
Decomposition, have been assessed on a perturbed turbulent velocity signal.
A numerical perturbation, whose parameters (amplitude and frequency) can
vary in time, has been added to the reference signal in order to mimic a
long-period flapping, i.e. the perturbation frequency remains lower than the
typical frequency featuring the energy-contained eddies of the turbulent flow.
The EMD algorithm has been used to recover, \textit{a posteriori},
both the reference signal and the perturbation. For that purpose, a rejection
procedure, acting such as a high-pass filter, has been designed. This procedure
is based on the "resemblance" criterion, introduced for the first time in this
paper, enabling to discriminate between the perturbed and the unperturbed
modes deduced from the EMD.

First, an extensive investigation of the EMD efficiency has been performed and
discussed in the case of a "mono-component" perturbation (sine wave) varying
both its amplitude and its frequency. It has been shown that the EMD performances
are mainly affected by the frequency parameter. This result has been emphasized
by the analysis of both the energy spectrum and the second-order structure function
revealing that the recovering procedure is significantly altered when the perturbation
frequency overlaps the frequency range of the energy-containing eddies. Although
the turbulent energy loss at large-scale, due to the high-pass filtering scheme,
the dynamics of both the inertial and the dissipative ranges
are successfully reproduced. Especially, we have pointed out that the energy
transfer through scales is satisfactorily recovered by studying the third-order
structure functions. Then, the relevance of the EMD algorithm has been addressed
in the specific case of a non-stationary perturbation (linear chirp).
Our results have shown that the recovering procedure developed in this
study performs with great success over the broad range swept by the perturbation
frequency.

This study highlights the potential of the EMD algorithm in the study of turbulence
and its interplay with coherent structures. These encouraging results rely on the
ability to extract a local trend, i.e. the local mean-value, in a fluctuating
signal which is of great interest in the framework of closed-loop flow control.

\bibliographystyle{unsrt}
\bibliography{MF_EMD_Arxiv_Nov2010}

\end{document}